\documentclass[reprint,prl,twocolumn,superscriptaddress,showpacs,showkeys,floatfix,preprintnumbers]{revtex4-1}

\usepackage{braket}
\usepackage{ulem}
\usepackage{xargs}
\usepackage{diagbox}
\usepackage{mathtools}
\usepackage[english]{babel}
\usepackage[utf8]{inputenc}
\usepackage{amsmath,amssymb,braket,slashed,mathrsfs}
\usepackage{pifont}
\usepackage{MnSymbol}
\usepackage{graphicx}
\usepackage{float}
\usepackage{subfigure}
\graphicspath{{./fig/}}
\usepackage{tikz}
\usetikzlibrary{shapes,arrows,positioning}
\tikzset{decision/.style={diamond, draw, fill=blue!20, text width=4.5em, text badly centered, inner sep=0pt}}
\tikzset{block/.style={rectangle, draw, fill=blue!20, text width=10em, text centered, rounded corners, minimum width=3.5cm}}
\tikzset{block1/.style={rectangle, draw, fill=blue!20, text width=18.5em, text centered, rounded corners, minimum width=3.5cm}}
\tikzset{line/.style={draw, -latex, thick}}
\usepackage{color}
\usepackage{hyperref}
\usepackage{multirow,array}

\usepackage{cleveref}
\begin{document}

\title{The anomalous long lifetime  of $^{14}$C revealed by \textit{ab\  initio} nuclear lattice EFT}

\author{Teng~Wang}\email{tenggeer@pku.edu.cn}\affiliation{School of Physics, Peking University, Beijing 100871, China.}
\author{Serdar~Elhatisari}\affiliation{Faculty of Natural Sciences and Engineering, Gaziantep Islam Science and Technology University,
  Gaziantep 27010, Turkey.}
\author{Xu~Feng}\email{xu.feng@pku.edu.cn}\affiliation{School of Physics, Peking University, Beijing 100871,
China.}\affiliation{Center for High Energy Physics, Peking University, Beijing 100871, China.}\affiliation{Collaborative Innovation Center of Quantum Matter, Beijing 100871, China.}\affiliation{Southern Center for Nuclear-Science Theory (SCNT), Institute of Modern Physics, Chinese Academy of Sciences, Guangdong 516000, China.}
\author{Bing-Nan~Lu}\email{bnlv@gscaep.ac.cn}\affiliation{Graduate School of China Academy of Engineering Physics, Beijing 100193, China.}
\author{Ulf-G.~Mei\ss ner}\email{meissner@hiskp.uni-bonn.de}\affiliation{Helmholtz-Institut f\"ur Strahlen- und Kernphysik, Bethe Center for Theoretical Physics and Cluster of Excellence Color meets Flavor, Universit\"at Bonn, D-53115 Bonn, Germany.}\affiliation{Institute for Advanced Simulation (IAS-4), Forschungszentrum J\"ulich, D-52425 J\"ulich, Germany.}\affiliation{Peng Huanwu Collaborative Center for Research and Education, International Institute for Interdisciplinary
  and Frontiers, Beihang University, Beijing 100191, China.}

\begin{abstract}

The 5730-year half-life of $^{14}$C, the physical basis of radiocarbon dating, is anomalously long compared to typical nuclear-physics expectations. Its origin has remained a subject of debate for many decades. Here we report an \textit{ab initio} nuclear lattice effective field theory (NLEFT) calculation of $^{14}$C $\beta$ decay. Despite the formidable challenge caused by Monte Carlo sign problems, using systematically optimized interactions and transition operators consistently derived from chiral effective field theory, we attack it through  a newly-developed multi-channel variational method and obtain a result consistent with the Gamow-Teller matrix element $M_\text{GT}^\text{exp}\simeq 2\times 10^{-3}$ measured with the current uncertainty of $O(10^{-2})$. This indicates a strong suppression compared to the 
simple order one expectation. We first show that chiral interactions and weak currents beyond leading-order are essential for quenching the GT matrix element to the physical value, among which the optimization of three-nucleon forces is indispensable.  We then illustrate that the quenching is deeply rooted in the ground-state structure of $^{14}$N as found in the nuclear shell model, where the competition between $S$- and $D$-wave components exists, sensitive to the interaction employed. The physical $^{14}$N ground state is found to be dominated by $D$-wave configurations, which constitutes the key factor for the quenching. The sensitivity of the decay matrix element to the fine-tuning of low-energy-constants is explored, revealing the prominent role of the $^3S_1$-channel two-nucleon contact force and the one-pion-exchange three-nucleon force.
This work eliminates the gap between shell-model and \textit{ab initio} studies of $^{14}$C $\beta$ decay,  provides a valid and straightforward explanation for the anomalous long lifetime of $^{14}$C, and turns
NLEFT into a practical tool for the systematic study of nuclear transitions.

\end{abstract}

\flushbottom
\maketitle

\thispagestyle{empty}

\section{Introduction}

The 5730-year half-life of $^{14}$C forms the physical backbone of radiocarbon
dating~\cite{Ajzenberg:1991,Chou:1993}, a cornerstone of modern archaeology, climatology, and ecology.
Yet this remarkable longevity is itself a nuclear puzzle:
the $\beta$ decay
$^{14}\mathrm{C}(0^+_1)\to{}^{14}\mathrm{N}(1^+_1)$ belongs to the allowed
Gamow-Teller (GT) class, one of the fastest and most common $\beta$-decay mechanisms, whose matrix element $M_\mathrm{GT}$ is expected to be
of order unity. However, the experimental value of this process, $M_{\mathrm{GT}}^{\mathrm{exp}}=2\times 10^{-3}$~\cite{Chou:1993},  is $2\sim3$ orders of magnitude smaller than the majority of allowed $\beta$-decay processes, corresponding to a transition strength (characterized by $\mathrm{log}\ ft$ value) $10^3\sim10^5$ times weaker than usual~\cite{Singh:1998}, making it the most hindered allowed transition~\cite{Kutschera:2019}. Historically, the puzzle has attracted broad and persistent interest~\cite{Talmi:2022, Kutschera:2019}, and extensive studies exist in the framework of the  nuclear shell model  ~\cite{Inglis:1953, Jancovici:1954, Zamick:1995,Fayache:1999, Suzuki:2003, Dai:2021, Fayache:1999, Aroua:2003, Holt:2009}.
It was originally proven that the nearly vanishing value
of $M_{\mathrm{GT}}$ cannot be explained if only central and spin-orbit forces are included~\cite{Inglis:1953},
and Ref.~\cite{Jancovici:1954} overcame this issue by additionally incorporating  tensor forces into the shell
model calculation. The role of the tensor force on $^{14}$C $\beta$ decay is further investigated in
Refs.~\cite{Zamick:1995,Fayache:1999, Suzuki:2003, Dai:2021}, while the sensitivity of $M_{\mathrm{GT}}$ to the
fine-tuning of both spin-orbit and tensor  forces is stressed in Ref.~\cite{Fayache:1999, Aroua:2003}. Recently, nuclear $ab \ initio$ calculations based on chiral effective field theory ($\chi$EFT) provide exact many-body wave functions starting from the fundamental symmetries of quantum chromodynamics, thus offering the most reductionist opportunity to investigate such abnormally fine-tuned phenomena in a model-independent manner~\cite{Holt:2009uk,Maris:2011, Ekstrom:2014}. Among them,
chiral three-nucleon (3N) forces are found to be essential for describing the large quenching
of $M_{\mathrm{GT}}$, while the effect of  two-body weak currents is
studied in Ref.~\cite{Ekstrom:2014}. Having identified the  correlations between many-body forces and the matrix element , however, the physical picture remains largely opaque~\cite{Talmi:2022}. In particular, it is unclear why 3N forces have a significant impact on  $M_{\mathrm{GT}}$, and the underlying mechanism connecting such a suppression of $\beta$ decay strength to the fine-structure of the complex nuclear wave functions awaits a detailed examination.  

To establish a firm microscopic origin of this anomaly,  systematic quantitative  discussions on nuclear structures using  state-of-the-art many-body algorithms are required, and the following shell-model ansatz could be of particular relevance~\cite{Jancovici:1954, Talmi:2022}: the ground state of $^{14}$C and $^{14}$N can be viewed as an
$^{16}$O-core plus two holes in the $p$-shell, whose orbital angular momentum can be coupled to $L=0,1$
for the former and $L=0, 1, 2$ for the latter. Since the leading order (LO) GT transition operator does not
change $L$, the decay would be greatly suppressed if the ground state of $^{14}$N carries exclusively $L=2$
components, thus providing a simple but elegant phenomenological explanation for $^{14}$C's longevity.  While the assumption of $^{14}$N's $D$-wave dominance has been  validated experimentally~\cite{Negret:2006}, the underlying microscopic mechanism is not understood yet, leaving a gap between the shell model and $ab\ initio$ methods.

In this work, we perform systematic studies on $^{14}$C $\beta$ decay through state-of-the-art nuclear lattice effective field theory
simulation~\cite{Dean:2009,Lahde:2019,Dean:2025}, based on previous works on $^{3}$H and $^{6}$He
$\beta$ decays~\cite{Elhatisari:2024,Teng:2025}. By employing systematically improved interactions and currents
derived from $\chi$EFT, we reveal the close relationship between the lifetime of $^{14}$C  and the inner structure of
$^{14}$N, validating the shell-model interpretation of $^{14}$C $\beta$ decay from $ab \ initio$ NLEFT calculation. Among different nuclear force components, we 
identify key interaction terms, including the one-pion-exchange 3N force, that critically influence $^{14}$N's structure, thus elucidating the sensitivity of $^{14}$C's lifetime to 3N forces.    
These help eliminate the theoretical gap between shell model and $ab\ initio$ works, and push forward the frontier in understanding the anomalous lifetime of $^{14}$C.

\section*{Lattice setup}

We employ  the high-fidelity next-to-next-to-next-to-leading order (N$^3$LO) lattice chiral interaction developed in
Ref.~\cite{Elhatisari:2022}, with lattice spacing $a=1.32$~fm. Two different sets of 3N forces are adopted
in the calculation: the first set is proposed and used in Ref.~\cite{Elhatisari:2022}, fitted only to
nuclear energies over a large range of nuclei, while the second set is simultaneously fitted to the binding energies
of selected light nuclei up to $^{16}$O, the magnetic dipole moment $\mu$ of $^{14}$N($1^+_1$) and the energy
difference $\Delta E$ between $^{14}$N(1$^+_1$) and $^{14}$N(1$^+_2$). For clarity, we label the former 3N
force as $V_{\mathrm{3N}}^{\mathrm{wfm}}$
and the latter as $V_{\mathrm{3N}}^{\mathrm{opt}}$, with the corresponding
Hamiltonians  as $H_{\chi}^{3\mathrm{N}_\mathrm{wfm}}$ and $H_{\chi}^{3\mathrm{N}_\mathrm{opt}}$, respectively.
The reason for including the
latter two observables when fitting $V_{\mathrm{3N}}^{\mathrm{opt}}$ is to optimize the $p$-shell structure of
the $A=14$ multiplets, as the
magnetic moment is a sensitive probe for  valence-nucleon
distributions~\cite{Schmidt:1937, Arima:1954, Li:2018}, while $\Delta E$ is shown to be
correlated with the lifetime of $^{14}$C~\cite{Ekstrom:2014}. We stress that fitting to $\Delta E$ provides a structural constraint on the $^{14}$N $p$-shell configuration mixing, but does not uniquely fix $M_\text{GT}$: as shown in the Supplementary Material, the quenching of $M_\text{GT}$ is jointly controlled by the $S$-$D$ wave composition of $^{14}$N and by the two-body weak currents, the latter being predicted by $\chi$EFT independently. To study the impact of higher-order interactions on $M_{\mathrm{GT}}$, we additionally consider the non-perturbative LO chiral Hamiltonian $H_S$~\cite{Elhatisari:2022}, which serves to solve the high-fidelity Hamiltonian perturbatively. For details on the lattice interactions and the determination of 3N forces, see Section S1 and S2 in Supplementary Material.

For the 
transition operator, we employ the LO GT operator,
\begin{equation}
    \label{eq:GT_opr_LO} O_{\mathrm{GT},\lambda}^{\mathrm{LO}} = -g_A\sum_{n=1}^A\sigma_{\lambda}(n)\tau_{+}(n)~, 
\end{equation}
with $g_A=1.275$ the free-space value of the nucleon axial-vector coupling constant measured in neutron
beta decay~\cite{ParticleDataGroup:2024}, $\sigma_{\lambda}$ the rank-one spherical tensor operator formed of spin Pauli matrices and
$\tau_{+}$ the isospin-raising Pauli matrix. We also incorporate the contribution of the
higher-order transition operator $O^{>\mathrm{LO}}_{\mathrm{GT},\lambda}$ induced by relativistic
corrections and two-body axial currents up to N$^3$LO~\cite{Baroni:2016, Krebs:2017, Krebs:2020}, which will be further discussed in Section S3 of  Supplementary Material.

\section*{Multi-channel variational method}

In nuclear lattice Monte Carlo simulations, a direct non-perturbative sampling of the full
Hamiltonian $H_\chi$ is prohibitive because of the severe sign problem~\cite{Troyer:2005} induced
by its complex operator structures. To alleviate the issue, the common strategy is to expand $|\Phi^0\rangle$, the ground state of $H_\chi$, around $|\Psi^0\rangle$, the ground state of the  simple Hamiltonian $H_S$ with milder sign problems, and to calculate observables order-by-order through  non-degenerate perturbation theory~\cite{Lu:2022, Liu:2025, Elhatisari:2022, Teng:2025, Ma:2023, Shen:2024, Ma:2024, Ren:2025}. For $M_{\mathrm{GT}}$ of $^{14}$C $\beta$ decay, however, we observe the failure of this approach: as will be fully discussed in the next section, the missing higher-order correlations make the valence structure of $|\Psi^0\rangle$ differ from $|\Phi^0\rangle$  significantly, yielding a LO result  $M^{\mathrm{LO}}_{\mathrm{GT}}\approx$ 2.4,  much larger than the experimental value. Consequently, if one expands $M_{\mathrm{GT}}$ around $|\Psi^0\rangle$ as a series of $H_\chi-H_S$,
it will not converge to zero until up to very high orders (if at all), which is impractical for numerical
Monte Carlo simulations.

In this work, we overcome this difficulty by constructing non-perturbative wave functions with optimized inner structures, based on the following multi-channel
variational method. We prepare several  shell-model trial states $|\Psi^i_{T,\mathrm{C}/\mathrm{N}}\rangle$
for $^{14}$C and $^{14}$N, respectively,  with the superscript $i$ denoting distinct valence-nucleon
distributions (see Section S4 of Supplementary Material for details). We evolve them through imaginary time projection using the simple Hamiltonian $H_S$,
\begin{equation}  |\Psi^i_{\mathrm{C}/\mathrm{N}}(\tau)\rangle = e^{-H_S\tau/2}|\Psi^i_{T,\mathrm{C}/\mathrm{N}}\rangle.
\end{equation}
For asymptotically large
projection time $\tau$,  the evolved states $|\Psi^i_{\mathrm{C/N}}(\tau)\rangle$, with $i=0, 1,\cdots$, form
a subspace composed of the low-lying eigenstates of $H_S$ for the initial and final nucleus.
We employ lattice Monte Carlo techniques to calculate their
inner products with each other,
\begin{equation}
\label{eq:N_ij}
  N^{ij}_{\mathrm{C/N}}(\tau) = \langle \Psi^{i}_{\mathrm{C/N}}(\tau)|\Psi^{j}_{\mathrm{C/N}}(\tau)\rangle,
\end{equation}
as well as their matrix elements with respect to the full Hamiltonian $H_\chi\in\{H_{\chi}^{3\mathrm{N}_{\mathrm{wfm}}}, H_{\chi}^{3\mathrm{N}_{\mathrm{opt}}}\}$,
\begin{equation}
\label{eq:H_ij}
H^{ij}_{\chi, \mathrm{C/N}}(\tau)=\langle \Psi^{i}_{\mathrm{C/N}}(\tau)|H_\chi|\Psi^{j}_{\mathrm{C/N}}(\tau)\rangle.
\end{equation}
The correlation functions in Eq.~(\ref{eq:N_ij}) and  (\ref{eq:H_ij}) allow for a variational determination
of the ground state  of $H_\chi$, denoted as $|\Phi^{0}_{\mathrm{C}/\mathrm{N}}\rangle$,
as a linear combination of $|\Psi^i_{\mathrm{C}/\mathrm{N}}(\tau)\rangle$, 
 \begin{equation}
\label{eq:eig_state_expansion}
|\Phi^0_{\mathrm{C}/\mathrm{N}}\rangle\propto v_{ \mathrm{C}/\mathrm{N}}^0|\Psi^0_{\mathrm{C}/\mathrm{N}}(\tau)\rangle
+v_{ \mathrm{C}/\mathrm{N}}^1|\Psi^1_{\mathrm{C}/\mathrm{N}}(\tau)\rangle+\cdots.
 \end{equation}
The coefficients $\boldsymbol{v}_{\mathrm{C}/\mathrm{N}}=(v_{ \mathrm{C}/\mathrm{N}}^0, v_{ \mathrm{C}/\mathrm{N}}^1,\cdots)^T$
can be solved from the following generalized eigenvalue equation,
 \begin{equation}
     \label{eq: GEVP}    
     H_{\chi}(\tau)\boldsymbol{v} = \varepsilon_0 N(\tau)\boldsymbol{v},
 \end{equation}
with $\varepsilon_0$ the smallest eigenvalue and   $N(\tau)$  and $H_{\chi}(\tau)$ a shorthand notation
of the matrices in Eqs.~(\ref{eq:N_ij}) and (\ref{eq:H_ij}). The GT matrix element $M_{\mathrm{GT}}$ can be
constructed from Eq.~(\ref{eq:eig_state_expansion}) as well as the following correlation function of the
transition operator $O_{\mathrm{GT},\lambda}$,
\begin{equation}
\label{eq: MGT_ij}
    M_{\mathrm{GT}, \lambda}^{ij}(\tau)= \langle \Psi_{\mathrm{N}}^i(\tau)|O_{\mathrm{GT},\lambda}|\Psi_{\mathrm{C}}^j(\tau)\rangle\, .
\end{equation}
The method to perform lattice Monte Carlo simulation of the correlation functions and the extraction of $M_{\mathrm{GT}}$ from them are left in Section S3 and S5 of Supplementary Material.

\begin{figure*}
\includegraphics[width=0.78\textwidth]{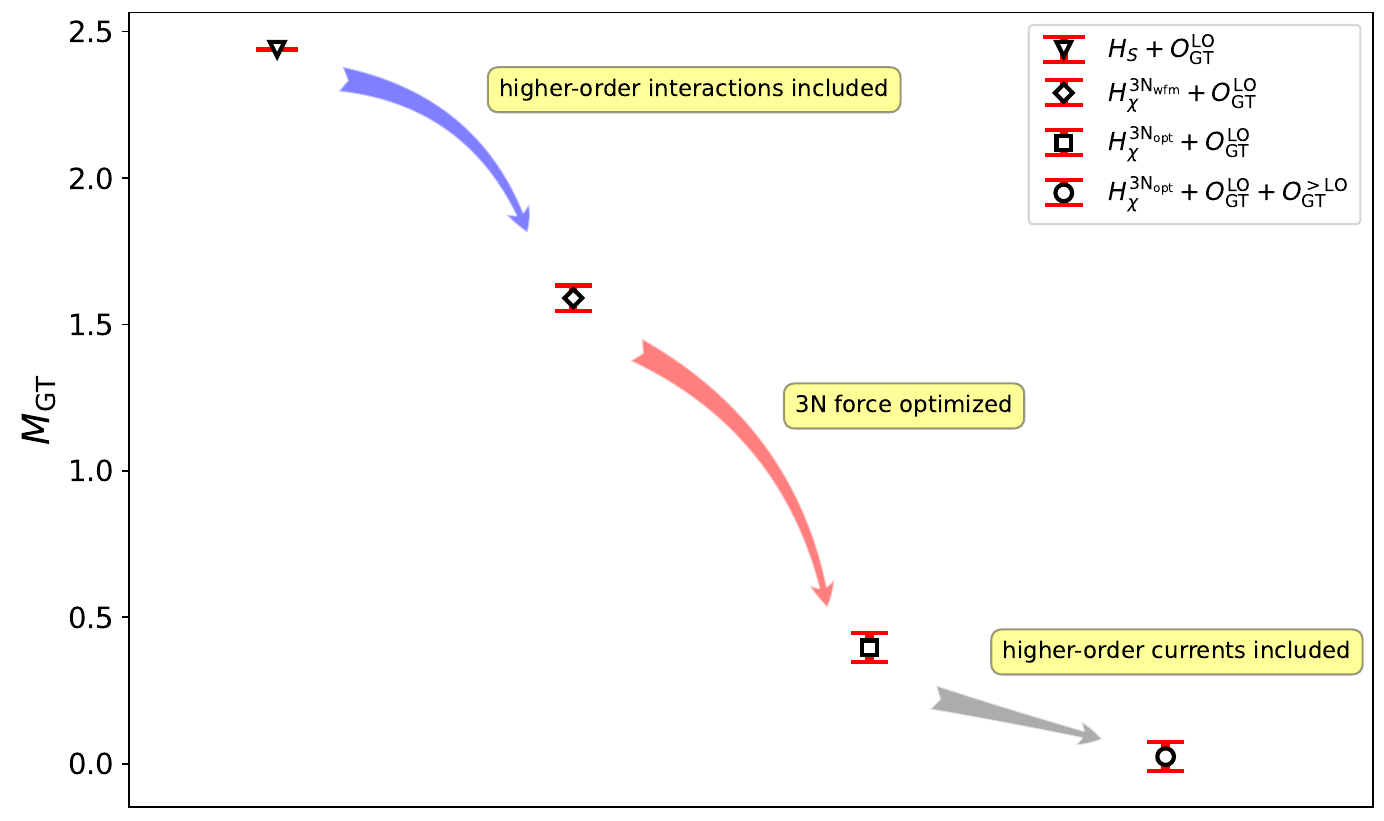}
\caption{The result of the GT matrix element of $^{14}$C $\beta$ decay, displayed in order with
  interactions and transition operators  optimized successively. The first point represents the result
  calculated from the LO Hamiltonian $H_S$ and the LO GT operator $O_{\mathrm{GT}}^\mathrm{LO}$. For the
  second point, $H_S$ is replaced by the high-fidelity N$^3$LO Hamiltonian $H_{\chi}^{3\mathrm{N}_{\mathrm{wfm}}}$. For the
  third point, $H_{\chi}^{3\mathrm{N}_{\mathrm{wfm}}}$ is replaced by $H_{\chi}^{3\mathrm{N}_{\mathrm{opt}}}$, whose 3N forces are improved. The last point additionally includes the correction from the higher-order GT operator $O^{>\mathrm{LO}}_{\mathrm{GT}}$.
  The statistical uncertainty is represented by the error bar.}
\label{fig:GT_vs_interaction}
\end{figure*}

\section*{Results and discussions}

In Figure~\ref{fig:GT_vs_interaction}, we show the results of  $M_{\mathrm{GT}}$, displayed  with the
interactions and transition operators  optimized successively. The prediction of the LO Hamiltonian
$H_S$ badly deviates from the experimental value $M_{\mathrm{GT}}^{\mathrm{exp}} = 2\times 10^{-3}$, due to the absence of
important spin-isospin correlations. Replacing $H_S$ with the original high-fidelity N$^3$LO Hamiltonian
$H_\chi^{3\mathrm{N}\mathrm{\mathrm{wfm}}}$ in Ref.~\cite{Elhatisari:2022}, $M_{\mathrm{GT}}$ is significantly quenched,
showing the need of including higher-order interactions. However, the result is still too large. It
is then greatly improved by $H_\chi^{3\mathrm{N}\mathrm{opt}}$, stressing the necessity of optimizing 3N forces.
Finally, after including the correction from higher-order weak currents, the full result $M_{\mathrm{GT}}=0.02(5)$ is statistically consistent with the experimental value $M_\text{GT}^\text{exp}\approx 2\times 10^{-3}$ with an uncertainty of $5\times 10^{-2}$, demonstrating that the combination of optimized 3N forces and higher-order weak currents brings the \textit{ab initio} prediction into the correct regime. We note that the large statistical uncertainty reflects the inherent challenge of sampling a nearly-vanishing observable with Monte Carlo methods, a further reduction of this uncertainty is a priority for future work

\begin{figure}[!htbp]
\centering
\includegraphics[width=0.48\textwidth]{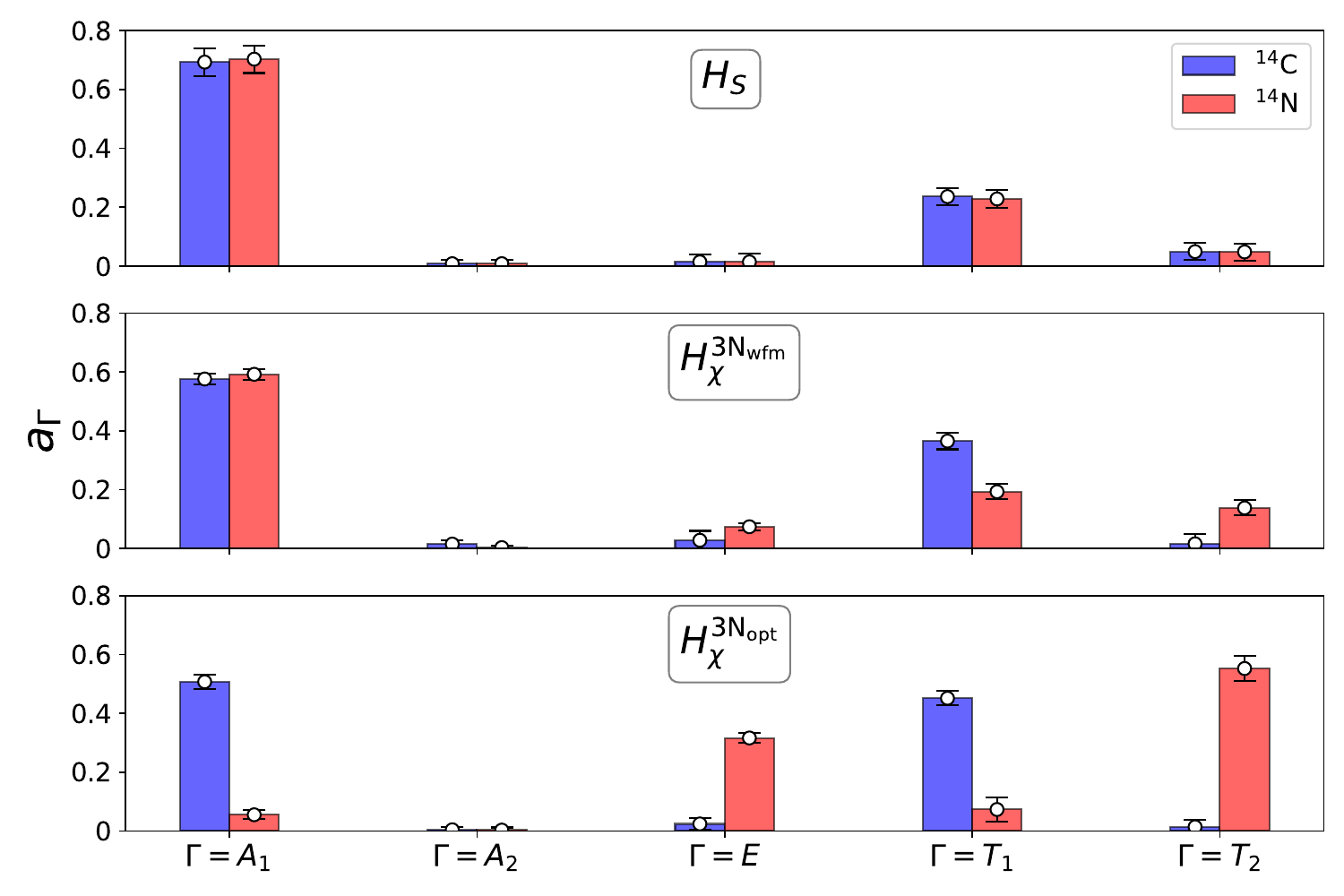}
\caption{The proportion of different cubic irrep components $a_{\Gamma}$ for the ground states of  $^{14}$C  and
$^{14}$N, represented by the blue and red bars respectively. The three panels from top to bottom denote the
results of $H_S$, $H_{\mathrm{\chi}}^{3\mathrm{N}_{\mathrm{wfm}}}$ and $H_{\mathrm{\chi}}^{3\mathrm{N}_{\mathrm{opt}}}$, in order. The statistical
uncertainty is represented by the error bar.  } \label{fig:irrep_vs_interaction}
\end{figure}

To understand the evolution of  $M_{\mathrm{GT}}$ in Figure~\ref{fig:GT_vs_interaction} and to further investigate
the origin of the longevity of $^{14}$C, we analyze the proportion
of different orbital angular-momentum components ($L$-components) of the initial and final nuclei, and we
explore how they change when different interactions are employed.
For lattice calculation, the proportion of the $L$-components can be easily identified using the
following quantity,
\begin{equation}
\label{eq:irrep_proj}
    a_\Gamma = \langle \Phi^{0}_H|P_\Gamma|\Phi_{H}^{0}\rangle, 
\end{equation}
where $|\Phi^0_H\rangle$ is the ground state of $^{14}$C or $^{14}$N for the Hamiltonian
$H\in\{H_0, H_\chi^{3\mathrm{N}_{\mathrm{wfm}}},  H_\chi^{3\mathrm{N}_{\mathrm{opt}}}\}$. $\Gamma\in \{A_1, A_2, E, T_1, T_2\}$
denotes the irreducible representation (irrep) of
the octahedral group $O$. The projection operator $P_{\Gamma}$ only acts on the orbital
wave function and projects it onto the irrep $\Gamma$ (see Section S8 of Supplementary Material for its explicit expression).  Eq.~(\ref{eq:irrep_proj}) can be viewed
as the lattice analog of the angular-momentum projection technique  in the continuum,
but is more convenient for the extraction of the $L$-components on the lattice,
using the decomposition rule of irreps of SO(3) into irreps of $O$ given in Table~\ref{tab:irrep_dec}.
In Figure~\ref{fig:irrep_vs_interaction},
we show the distribution of $a_\Gamma$ of $^{14}$C and $^{14}$N for different Hamiltonians. Starting
from $H_S$, the distributions of $^{14}$C and $^{14}$N are almost identical, both concentrated in the $A_1$ and $T_1$
irreps. This means that  the ground states of $H_S$ for both nuclei are  dominated by $L=0$ and $L=1$. This still
holds for $^{14}$C when the high-fidelity Hamiltonian is used, suggesting the stability of the $^{14}$C shell
structure against the change of interactions. In contrast, the structure of $^{14}$N exhibits greater
sensitivity to the interaction. After replacing $H_S$ with $H_{\chi}^{3\mathrm{N}_{\mathrm{wfm}}}$, there is a
noticeable increase of $E$- and $T_2$-contributions and a decrease of the $T_1$-contribution to the $^{14}$N
wave function, although the change is relatively mild. The significant change happens for $H_{\chi}^{3\mathrm{N}_{\mathrm{opt}}}$,
where $^{14}$N is strongly dominated by $E$ and $T_2$,  indicating that it is intensely populated by $L=2$ components.
The analysis above reveals the critical competition of $S$- and $D$-wave configurations in the $^{14}$N ground state and its
sensitivity to the interaction employed. The calculation using the optimized Hamiltonian $H_{\chi}^{3\mathrm{N}_{\mathrm{opt}}}$
suggests that the physical ground state of $^{14}$N is dominated by the $D$-wave, which is the key factor for
the quenching of $M_{\mathrm{GT}}$. Our finding agrees with the shell-model stuides mentioned in the introduction
and confirms the experimental finding in Ref.~\cite{Negret:2006}. In Section S9 of Supplementary Material, we show the
duality between the NLEFT results and the shell-model scenario, detailing how chiral 2N and 3N forces balance
different shell-model configurations in $^{14}$N. We mention that although $^{14}$N is not entirely dominated by
the $D$-wave, the higher-order weak currents further quench $M_{\mathrm{GT}}$ and bring it into consistency
with  experiment.

\begin{table}
\resizebox{230pt}{!}{
		\begin{tabular}{c| c |c |c |c |c }
			\hline\hline
			      $L$ & 0 & 1& 2 & 3& 4   \\
                 \hline
                 $\Gamma$&$A_1$&$T_1$&$E\oplus T_2$ & $A_2\oplus T_1\oplus T_2$ & $A_1\oplus E\oplus T_1\oplus T_2$\\
		\hline\hline
		\end{tabular}}
\caption{Decompositions of orbital angular momentum $L\le 4$ into irreps $\Gamma$ of the
  octahedral group $O$.}
	\label{tab:irrep_dec}
\end{table}

Based on  Figure~\ref{fig:GT_vs_interaction} and Figure~\ref{fig:irrep_vs_interaction}, we discuss the
collapse of non-degenerate
perturbation theory and the necessity to use the multi-channel variational method in this work.
Figure~\ref{fig:irrep_vs_interaction} shows that the ground state of $H_S$, $|\Psi^0_{\mathrm{N}}\rangle$, is
dominated by the $S$-wave and has a distinct shell structure from the physical ground state, which explains the large deviation of the first data point from the experimental value in Figure~\ref{fig:GT_vs_interaction}. Expanding $M_{\mathrm{GT}}$ perturbatively around this `wrong’ wave function would either diverge or suffer from slow convergence. In comparison, the multi-channel variational method developed here is equivalent
to LO degenerate perturbative theory,  which  treats the ground state
 and low-lying excited states of $H_S$ on
 equal footing from the very beginning, which systematically
optimizes the valence structure and improves the convergence of the perturbative series. In Section S10 of Supplementary Material, we provide additional information to validate the improved perturbative convergence of this method.

\begin{figure*}
\centering
\includegraphics[height=1.8in]{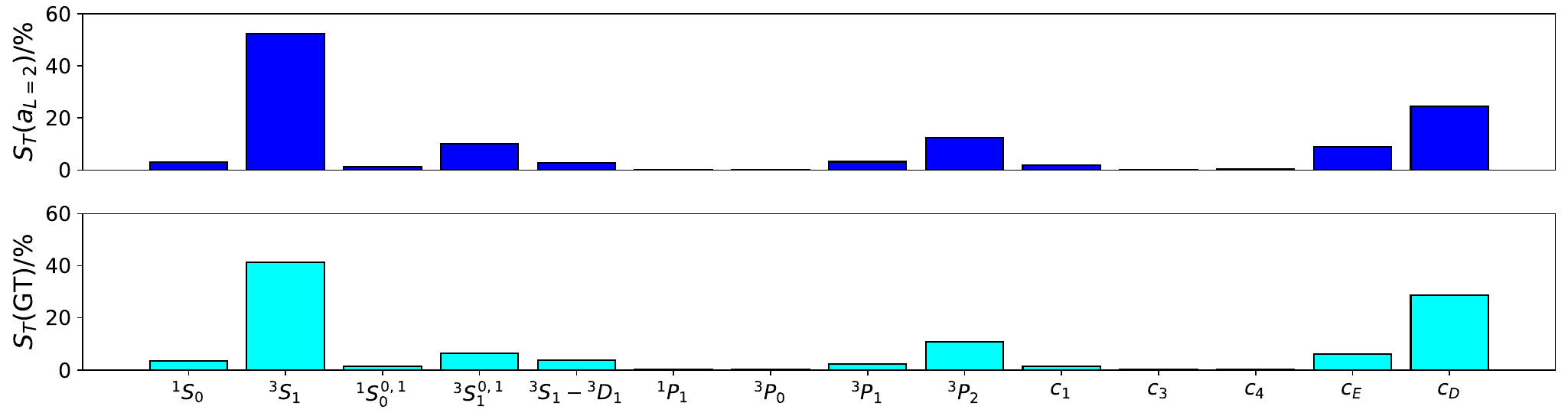}
\caption{ ({\bf Upper panel}) The total-order sensitivity
  indices $S_T$ of $a_{L=2}$ defined in Eq.~(\ref{eq:a_L2}) for different LECs in $H_\chi^{3\mathrm{N}_{\mathrm{opt}}}$.
  ({\bf Lower panel}) The total-order sensitivity
  indices $S_T$ of the GT matrix element of $^{14}$C $\beta^-$ decay for different LECs in
  $H_\chi^{3\mathrm{N}_{\mathrm{opt}}}$. The labels from $^{3}$S$_1$ to $^{3}$P$_2$ represent the 2N forces up to NLO.
  $c_1, c_3$ and $c_4$ are pion-nucleon couplings of the two-pion-exchange 3N force.
  $c_D$ and $c_E$ are the LECs of the 3N OPE force and the 3N contact term.}
\label{fig:ST_distribution}
\end{figure*}

To quantify how much the fine-tuning of different components of the nuclear force influences the
structure of $^{14}$N and the decay of $^{14}$C, the method of Sobol's
global sensitivity analysis (GSA)~\cite{Sobol:2001, Saltelli:2002, Saltelli:2010} recently
flourishing in nuclear $ab \ initio$ calculations provides a robust
tool~\cite{Ekstrom:2019, Belley:2024, Sun:2025, Becker:2026}.   In this work, we
apply it to both $M_{\mathrm{GT}}$ and the proportion of the $L=2$ component for $^{14}$N ground state,
\begin{equation}
\label{eq:a_L2}
    a_{L=2}= a_{\Gamma=E} + a_{\Gamma=T_2}.
\end{equation}
 We consider all 2N forces up to next-to-leading-order
(NLO) and 3N forces in $H_\chi^{3\mathrm{N   }_{\mathrm{opt}}}$, and we sample the corresponding low-energy-constants (LECs) 
 in a hypercube region bounded by $\pm 10\%$ around their central values. Since $H_\chi^{3\mathrm{N   }_{\mathrm{opt}}}$
 is treated perturbatively, the numerous
 model samples required by GSA can be worked out efficiently, and we leave the relevant technical details in Section S11 of Supplementary Material. To see which interaction term contributes
 the largest variance of the samples, we calculate
the total-order sensitivity
indices $S_{T}$ of each term for $a_{L=2}$ and $M_{\mathrm{GT}}$, respectively, and the corresponding
distribution of $S_{T}$ is displayed in Figure~\ref{fig:ST_distribution}.  Not surprisingly, we
find the upper and lower distributions highly similar to each other, demonstrating that $^{14}$C $\beta$ decay is
highly correlated with the $L=2$ component of $^{14}$N. We also
find that the variance of these two quantities can be mainly attributed to the  3N OPE term $V_{3\mathrm{N}}^{(c_D)}$,
consistent with previous \textit{ab initio} calculations~\cite{Maris:2011, Ekstrom:2014}, as well as the 2N
contact term of the $^3S_1$ partial wave, $V_{^3 S_1}$, which is identified for the first time. In comparison,
the influence of other terms are relatively small.

\begin{figure}[!htbp]
\includegraphics[width=0.48\textwidth]{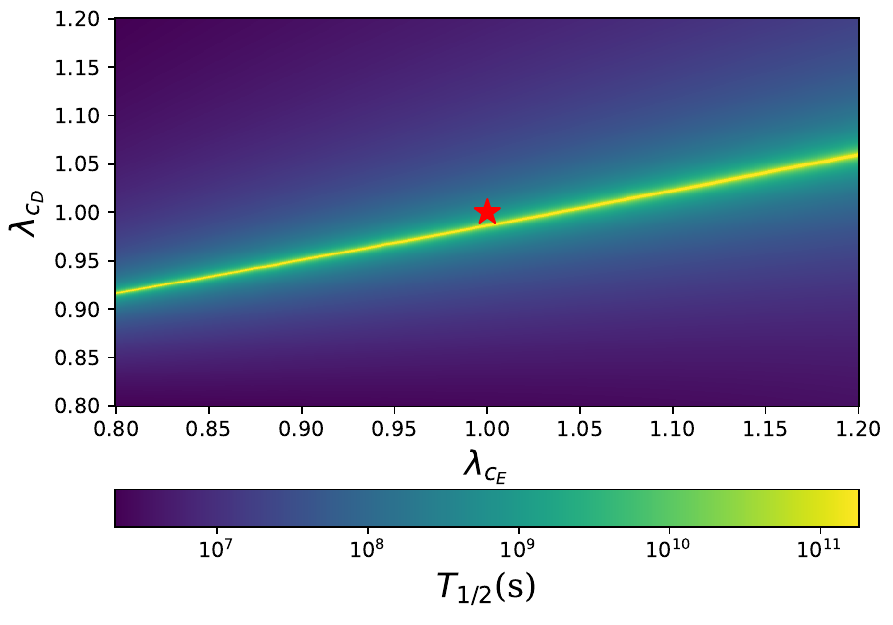}
\caption{The contour distribution of the half-life $T_{1/2}$ of $^{14}$C
versus the variation of parts of the 3N force.
  $\lambda_{c_D}$ and $\lambda_{c_E}$ are scaling factors multiplied with OPE and contact 3N terms, respectively.
  The red star  in the middle represents the optimized values of 3N forces used in $H_{\chi}^{3\mathrm{N}_{\mathrm{opt}}}$. The deviation from the experimental measured lifetime $T_{1/2}^{\mathrm{exp}}=1.807\times 10^{11}\mathrm{s}$ is because of the use of the central value $M_{\mathrm{GT}}=0.02$ in the calculation.}
\label{fig:C14_lifetime_contour}
\end{figure}

To better visualize the sensitivity of the $^{14}$C lifetime to the 3N forces, we vary the $c_D$ and $c_E$ terms
independently by multiplying their optimal values in $H_{\chi}^{3\mathrm{N}_{\mathrm{opt}}}$ with scaling factors $\lambda_{c_D}$
and $\lambda_{c_E}$, respectively. Using the formula  for the lifetime $T_{1/2}$ of $^{14}$C in  Supplementary Material,  we calculate it as a function of
$\lambda_{c_D}$ and $\lambda_{c_E}$ and show its contour plot in Figure~\ref{fig:C14_lifetime_contour}. The result corresponding to the central value $M_{\mathrm{GT}}=0.02$  of the optimized Hamiltonian $H_{\chi}^{3\mathrm{N}_{\mathrm{opt}}}$ is represented by the red star.  We find that the allowed region of the experimentally measured lifetime of $^{14}$C, i.e.
$T_{1/2}^{\mathrm{exp}}=1.807\times 10^{11}\mathrm{s}$~\cite{Chou:1993}, is constrained into a
narrow line. Any small variation of $\lambda_{c_D}$ and $\lambda_{c_E}$ perpendicular to this line
would drastically decrease the lifetime by one or more orders.  This observation  suggests the $^{14}$C lifetime
as a useful tool for constraining the 3N force~\cite{Holt:2009}, a frontier for $ab \ initio$
nuclear many-body calculations~\cite{Hebeler:2020}.

\section*{Concluding remarks}

In this work, we attack the long-standing challenge in  predicting and
understanding $^{14}$C $\beta$ decay in the framework of nuclear lattice effective
field theory. By performing a systematic analysis of the interplay between nuclear interactions,
nuclear structure and the transition matrix element, we link the shell-model and \textit{ab initio}
interpretation of $^{14}$C $\beta$ decay in a valid and intuitive manner, thus elucidating the long-standing
puzzle of the  anomalous long lifetime of $^{14}$C. We note that the current Monte Carlo statistical uncertainty on $M_\text{GT}$ is large, and a systematic quantification of chiral truncation and lattice-spacing errors remains for future work; nevertheless, the qualitative conclusion that $D$-wave dominance in $^{14}$N is the key quenching mechanism is robust.  Quantitative studies along this line with controlled theoretical uncertainties could undoubtedly promote the
understanding of the chiral interaction in the nuclear many-body environment, potentially improving
its prediction for other important electroweak processes. 

A major obstacle constraining the application of NLEFT is the Monte Carlo sign problem. Though
non-degenerate perturbation theory has been commonly employed for its mitigation, leading to much progress,
we point out that it may suffer from slow convergence for some observables such as $^{14}$C $\beta$ decay.
The multi-channel variational method proposed here helps to reorganize the perturbative series
and enhance its convergence,  providing new insights for the perturbative alleviation of the
sign problem. As the range of applicability of NLEFT has been extended to heavy
nuclei recently~\cite{Niu:2025, Hildenbrand:2025}, this method can be  directly used to study
$\beta$ decay processes key to $r$-process
nucleosynthesis~\cite{Mumpower:2015, Kajino:2019, Cowan:2019, Arcones:2022}. Combined
with the multi-reference trial state developed in Ref.~\cite{Wang:2025_MR}, it also allows
a simultaneous precise determination of the nuclear ground
and the excited states of the
high-fidelity chiral Hamiltonian, opening up the possibility for systematic studies of
configuration mixings and deformations through NLEFT.

\medskip


\section{acknowledgments}
 X.F. and T.W. were supported in part by NSFC of China under Grants No. 12125501 and No. 12550007.
B.N.L. was supported by NSAF No. U2330401 and National Natural Science Foundation of China with Grant
Nos. 12275259, 12547105.
The work of U.G.M. was supported in part by the Deutsche Forschungsgemeinschaft (DFG, German Research Foundation) under Germany's
Excellence Strategy – EXC 3107 – Project-ID~533766364, by the European Research Council (ERC) under the European Union's Horizon 2020 research and innovation programme (EXOTIC, grant agreement No. 101018170) and by the CAS President's International Fellowship Initiative (PIFI) (Grant No.~2025PD0022). S.E. was supported in part by Scientific and Technological Research Council
of Turkey (TUBITAK project no. 123F464).

\bibliography{bib_C14}

\clearpage

\setcounter{page}{1}
\renewcommand{\thepage}{Supplementary Material -- S\arabic{page}}
\setcounter{table}{0}
\renewcommand{\thetable}{S\,\arabic{table}}
\setcounter{equation}{0}
\renewcommand{\theequation}{S\,\arabic{equation}}
\setcounter{figure}{0}
\renewcommand{\thefigure}{S\,\arabic{figure}}

\begin{appendix}
\begin{onecolumngrid}

\section*{Supplementary material}

\subsection{S1. Nuclear lattice effective field theory}

\label{sec:S1}

Nuclear lattice effective field theory (NLEFT) is an $ab\ initio$ nuclear many-body approach that combines the framework of effective field
theory (EFT) with stochastic Monte Carlo techniques~\cite{Dean:2009,Lahde:2019}. In NLEFT calculations, the four-dimensional Euclidean spacetime is discretized into discontinuous lattice points confined in a finite cubic box. The  box size is $L^3\times L_t$, with $L$ and $L_t$ the cubic length along the spatial and temporal direction, respectively.  The spatial lattice size $a$ is usually taken as $1$-$2$~fm to regularize the high-momentum physics.  Nucleons are constrained onto the lattice sites with interactions among them  discretized properly. To deal with complex nuclear many-body correlations, the method of auxiliary transformation is employed to decompose two- and three-body potentials into couplings between auxiliary fields and single nucleons. The fluctuation of the auxiliary fields is then simulated through lattice Monte Carlo technique, through which one can calculate nuclear correlation functions and extract the observables of interest.

On the one hand, NLEFT is deeply rooted in low-energy QCD, as the high-fidelity chiral interaction originating from the spontaneous breaking of chiral symmetry of QCD is employed for accurate lattice simulations.  On the other hand, the powerful lattice Monte Carlo technique not only allows to treat all nucleons as dynamical degrees of freedom, which captures the essential many-body correlations, but also leads to the mild power-law scaling dependence  of the computational cost on the nuclear mass number. Due to the above merits, NLEFT has matured into a leading framework for the investigation
of nuclear systems, and its range of applicability has been extended from light to medium-mass and heavy nuclei~\cite{Niu:2025, Hildenbrand:2025}.

We employ  the high-fidelity N$^3$LO lattice chiral interaction together with the wave function matching method developed in
Ref.~\cite{Elhatisari:2022}, which allows for the precise determination of nuclear binding energies. The wave function matching method is an efficient approach to alleviating sign problems, which unitarily transforms the original Hamiltonian into the new high-fidelity Hamiltonian $H_\chi$, whose
wave function matches those of a computationally simpler Hamiltonian $H_S$ at short distances.  This construction accelerates the convergence of the perturbative expansion in $H_{\chi} - H_S$. In the following, we introduce details on the expressions of $H_S$ and $H_\chi$. For definiteness, we define the following locally smeared SU(4)-symmetric nucleon density operator,
\begin{equation}
\label{eq:rho_d}
\begin{aligned}
    \rho^{(d)}(\boldsymbol{n}) &= \sum_{i=0}^1\sum_{j=0}^1 \Big{[}a^\dagger_{i,j}(\boldsymbol{n})a_{i,j}(\boldsymbol{n})+s_\mathrm{L} \sum_{|\boldsymbol{n}-\boldsymbol{n}'|=1}^d a^\dagger_{i,j}(\boldsymbol{n}')a_{i,j}(\boldsymbol{n}')\Big{]},
\end{aligned}
\end{equation}
and the locally smeared spin-isospin dependent nucleon density operator,
\begin{equation}
\label{eq:rho_SI_d}
    \begin{aligned}      \rho^{(d)}_{S,I}(\boldsymbol{n}) &= \sum_{i,i'=0}^1\sum_{j,j'=0}^1 \Big{[}a^\dagger_{i,j}(\boldsymbol{n})[\sigma_S]_{i,i'}[\tau_I]_{j,j'}a_{i',j'}(\boldsymbol{n})+s_\mathrm{L} \sum_{|\boldsymbol{n}-\boldsymbol{n}'|=1}^d a^\dagger_{i,j}(\boldsymbol{n}')[\sigma_S]_{i,i'}[\tau_I]_{j,j'}a_{i',j'}(\boldsymbol{n}')\Big{]}.\\
    \end{aligned}
\end{equation}
In the above, $a^{\dagger}_{ij}(\boldsymbol{n})$ and $a(\boldsymbol{n})$ are nucleon creation and annihilation operators, respectively. $i=0,1$ and $j=0,1$ are spin and isospin indices, respectively.  $\sigma_S$ and $\tau_I$ are spin and isospin Pauli matrices, with $S, I=1, 2, 3$. $s_{\mathrm{L}} =0.07$ is the smearing parameter controlling the strength of locality, while $d$ controls the range of local smearing.

The  LO Hamiltonian $H_S$ consists of the following four terms,
\begin{equation}
\label{eq:H_S_expression}
    H_S = K + V_{\mathrm{SU}(4)}+ V_{I}+V^{\Lambda_\pi = 180\mathrm{MeV}}_{\mathrm{OPE}},
\end{equation}
with $K$  the kinetic energy term, $V_{\mathrm{SU}(4)}$  the SU(4)-symmetric LO contact term, $V_I$ the isospin-dependent LO contact term and $V^{\Lambda_\pi}_{\mathrm{OPE}}$ the OPE term. $\Lambda_\pi$ is the momentum-space cutoff for regulating the short-range singularity in the OPE potential. For the explicit expression of the individual terms, we refer the reader to Ref.~\cite{Elhatisari:2022}. Note that $V_{\mathrm{SU}(4)}$ is much stronger than $V_I$ and $V^{\Lambda_\pi}_{\mathrm{OPE}}$, so $H_S$ is approximately SU(4)-symmetric.

The interactions in the N$^3$LO Hamiltonian $H_\chi$ can be separated into 2N and 3N parts. Since the key difference in this paper from Ref.~\cite{Elhatisari:2022} is the refit of 3N forces (collectively denoted as $V_{\mathrm{3N}}$), we focus on $V_{\mathrm{3N}}$ below and refer the reader to Ref.~\cite{Elhatisari:2022} for details about remaining interactions. $V_{\mathrm{3N}}$ consists of all 3N terms at next-to-next-to-leading-order (N$^2$LO), including the contact potential $V_{\mathrm{3N}}^{(c_E)}$, the one-pion exchange potential $V_{\mathrm{3N}}^{(c_D)}$, and the two-pion exchange (TPE)
potential $V_{\mathrm{3N}}^{(\mathrm{TPE})}$,
\begin{equation}
\label{eq:V_3N_in_short}
    V_{\mathrm{3N}} = V_{\mathrm{3N}}^{(\mathrm{TPE})} + V_{\mathrm{3N}}^{(c_D)} +  V_{\mathrm{3N}}^{(c_E)}.
\end{equation}
Among them, the TPE potential $V_{\mathrm{3N}}^{(\mathrm{TPE})}$ can be separated into three parts,
\begin{equation}
 \begin{aligned}  V_{\mathrm{3N}}^{(\mathrm{TPE1})}
 =&\frac{c_3 g_A^2}{4 F_\pi^4}\sum_I \sum_{S,S',S''}\sum_{\boldsymbol{n},\boldsymbol{n}',\boldsymbol{n}''}:\rho^{(0)}_{S',I}(\boldsymbol{n}')\rho^{(0)}_{S'', I}(\boldsymbol{n}'')\rho^{(0)}(\boldsymbol{n}): f_{S',S}(\boldsymbol{n}'-\boldsymbol{n})f_{S'',S}(\boldsymbol{n}''-\boldsymbol{n}),\\
 V_{\mathrm{3N}}^{(\mathrm{TPE2})} 
 =&-\frac{c_1 g_A^2 M_\pi^2 }{2 F_\pi^4}\sum_I \sum_{S',S''}\sum_{\boldsymbol{n},\boldsymbol{n}',\boldsymbol{n}''}:\rho^{(0)}_{S',I}(\boldsymbol{n}')\rho^{(0)}_{S'', I}(\boldsymbol{n}'')\rho^{(0)}(\boldsymbol{n}):
 f_{S'}(\boldsymbol{n}'-\boldsymbol{n}) f_{S''}(\boldsymbol{n}''-\boldsymbol{n}),\\
 V_{\mathrm{3N}}^{(\mathrm{TPE3})}
 =&-\frac{c_4 g_A^2 }{8 F_\pi^4}\sum_{I_1,I_2,I_3}\sum_{S_1, S_2, S_3}\sum_{S',S''}\sum_{\boldsymbol{n},\boldsymbol{n}',\boldsymbol{n}''}\varepsilon_{I_1,I_2,I_3}\varepsilon_{S_1, S_2, S_3}
 :\rho^{(0)}_{S',I_1}(\boldsymbol{n}')\rho^{(0)}_{S'', I_2}(\boldsymbol{n}'')\rho^{(0)}_{S_3, I_3}(\boldsymbol{n}): \\
 &f_{S',S_1}(\boldsymbol{n}'-\boldsymbol{n})
f_{S'',S_2}(\boldsymbol{n}''-\boldsymbol{n}).
 \end{aligned} 
\end{equation}
In the above expressions, the density operators defined in Eq.~(\ref{eq:rho_d}) and~(\ref{eq:rho_SI_d}) have been used.  $g_A=1.287$ is the axial-vector coupling constant adjusted to account for the Goldberger-Treiman discrepancy~\cite{Fettes:1998}, $F_\pi=92.2\,\mathrm{MeV}$ is the pion decay constant, $c_1 = -1.10\,\mathrm{GeV}^{-1}, c_3 = -5.54\,\mathrm{GeV}^{-1}$ and
$c_4 = 4.17\,\mathrm{GeV}^{-1}$ are dimension-two pion–nucleon coupling constants taken from a Roy-Steiner analysis of pion-nucleon scattering~\cite{Hoferichter:2015}. Also, $\epsilon_{i,j,k}$ is the Levi-Civita symbol and the colons :: indicate normal ordering. The functions $f_S$ are $f_{S, S'}$ are defined as
\begin{equation}
\begin{aligned}
   & f_{S}(\boldsymbol{n}-\boldsymbol{n}') = \frac{1}{L^3}\sum_{\boldsymbol{q}}\frac{q_S e^{-i\boldsymbol{q}\cdot(\boldsymbol{n}-\boldsymbol{n}')-(q^2+M_\pi^2)/\Lambda_\pi^2}}{q^2+M_\pi^2},\\
   &f_{S,S'}(\boldsymbol{n}-\boldsymbol{n}') = \frac{1}{L^3}\sum_{\boldsymbol{q}}\frac{q_{S'}q_S e^{-i\boldsymbol{q}\cdot(\boldsymbol{n}-\boldsymbol{n}')-(q^2+M_\pi^2)/\Lambda_\pi^2}}{q^2+M_\pi^2},
\end{aligned}
\end{equation}
with $M_\pi=134.98\,\mathrm{MeV}$ the pion mass and $\Lambda_\pi = 300\,\mathrm{MeV}$ the momentum space regulator. Following Ref.~\cite{Elhatisari:2022}, the contact potential $V_{\mathrm{3N}}^{(c_E)}$ and OPE potential $V_{\mathrm{3N}}^{(c_D)}$ are locally smeared with different values of $d$. Besides, two additional SU(4) symmetric terms accounting for different 3N configurations are included into $V_{\mathrm{3N}}^{(c_E)}$, which are denoted by $V^{(l)}_
{c_E}$ and $V^{(t)}_
{c_E}$. Therefore, the complete expressions of $V_{\mathrm{3N}}^{(c_D)}$ and $V_{\mathrm{3N}}^{(c_E)}$ are
\begin{equation}
\label{eq:V_cD_form}
    V_{\mathrm{3N}}^{(c_D)} = \sum_{d=0}^2 V_{c_D}^{(d)}, 
\end{equation}
and
\begin{equation}
\label{eq:V_cE_form}
    V_{\mathrm{3N}}^{(c_E)} =    \sum_{d=0}^2 V_{c_E}^{(d)}+ V^{(l)}_
{c_E} + V^{(t)}_
{c_E}.
\end{equation}
For $V_{c_D}^{(d)}$ and $V_{c_E}^{(d)}$, their full expressions are
\begin{equation}
\label{eq:cDcE_term}
\begin{aligned}
   & V_{c_D}^{(d)} = -\frac{c_D^{(d)}g_A}{4 F_\pi^4 \Lambda_\chi}\sum_{\boldsymbol{n},S,I}\sum_{\boldsymbol{n}', S'} :\rho_{S',I}^{(d)}(\boldsymbol{n}')\rho_{S,I}^{(d)}(\boldsymbol{n})\rho^{(d)}(\boldsymbol{n}): 
    f_{S',S}(\boldsymbol{n}'-\boldsymbol{n}),\\
    &V_{c_E}^{(d)} = \frac{1}{6}\frac{c_E^{(d)}}{2 F_\pi^4 \Lambda_\chi}\sum_{\boldsymbol{n}}:\Big{[}\rho^{(d)}(\boldsymbol{n})\Big{]}^3:,
\end{aligned}
\end{equation}
with $\Lambda_\chi = 700\,\mathrm{MeV}$. $c_D^{(d)}$ and $c_E^{(d)}$ are unknown LECs for fit. $V^{(l)}_{c_E}$ and $V^{(t)}_{c_E}$ are SU(4) symmetric terms accounting for 3N configurations of prolate shape and oblate shape, respectively,
\begin{equation}
\begin{aligned}
    V^{(l)}_{c_E} =& c_E^{(l)}\sum_{\boldsymbol{n},\boldsymbol{n}',\boldsymbol{n}'' }:\rho^{(0)}(\boldsymbol{n})\rho^{(0)}(\boldsymbol{n}')\rho^{(0)}(\boldsymbol{n}''):\delta_{|\boldsymbol{n}-\boldsymbol{n}'|,1} \ \delta_{|\boldsymbol{n}-\boldsymbol{n}''|,1}\ \delta_{|\boldsymbol{n'}-\boldsymbol{n}''|,2},\\
    V^{(t)}_{c_E} =& c_E^{(t)}\sum_{\boldsymbol{n},\boldsymbol{n}',\boldsymbol{n}'' }:\rho^{(0)}(\boldsymbol{n})\rho^{(0)}(\boldsymbol{n}')\rho^{(0)}(\boldsymbol{n}''):
    \delta_{|\boldsymbol{n}-\boldsymbol{n}'|,\sqrt{2}}\  \delta_{|\boldsymbol{n}-\boldsymbol{n}''|,\sqrt{2}}\ \delta_{|\boldsymbol{n'}-\boldsymbol{n}''|,\sqrt{2}}.
\end{aligned} 
\end{equation}
The $c_{E}^{(l)}$ and $c_E^{(t)}$ are also LECs to be fitted.

\subsection{S2. Determination of the 3N forces}

\label{sec:S2}

We perform a combined fit of the eight 3N LECs, i.e. $c_D^{(0)}, c_D^{(1)}, c_D^{(2)}, c_E^{(0)}, c_E^{(1)}, c_E^{(2)}, c_E^{(L)}$ and $c_E^{(t)}$, to the following quantities:  1. the energies of selected light nuclei up to $^{16}$O;
2. the  magnetic dipole moment $\mu=0.404\,\mu_N $ of  the $^{14}$N ground state;
3. the energy gap $\Delta E= 3.948\,\mathrm{MeV}$ between the $^{14}\mathrm{N}(1^+_1)$ and $^{14}\mathrm{N}(1^+_2)$ states. The reason to choose these for the fit has been explained in the main text. When fitting 3N forces, the 2N LECs are fixed to their optimal values~\cite{Elhatisari:2022}, and we determine the 3N LECs by minimizing
\begin{equation}
    \chi^2 = \sum_{i}\frac{(E_{\mathrm{latt}}^{(i)}-E_{\mathrm{exp}}^{(i)})^2}{\epsilon_i^2}+\frac{(\mu_{\mathrm{latt}}-\mu_{\mathrm{exp}})^2}{\epsilon_\mu^2}+\frac{(\Delta E_{\mathrm{latt}}-\Delta E_{\mathrm{exp}})^2}{\epsilon_{\Delta E}^2},
\end{equation}
where $i$ denotes the  nuclei whose energies are used for fit. $O_{\mathrm{exp}}$ and $O_{\mathrm{latt}}$
denote the experimental value and the NLEFT prediction, respectively. The statistical uncertainties for $\mu$ and $\Delta E$ are sufficiently large, so we use them as $\epsilon_{\mu}$ and $\epsilon_{\Delta E}$. The uncertainty $\epsilon_i$ for the energy consists of two parts,
\begin{equation*}
    \epsilon_i^2 = \epsilon_{i, \mathrm{sta}}^2+\epsilon_{i, \chi}^2,
\end{equation*}
where $\epsilon_{i, \mathrm{sta}}$ denotes the statistical error and $\epsilon_{i, \chi}$ represents the uncertainty induced by the truncation of the chiral nuclear forces. We include the latter because  $\epsilon_{i,\mathrm{sta}}$ of few-body nuclei such as $^{4}$He is too small, for which the truncation uncertainty is non-negligible. We estimate $\epsilon_{i, \chi}$ as the variance of 2N energy caused by  the variation of 2N forces, which is calculated from the  configurations of 2N LECs generated in Ref.~\cite{Elhatisari:2022}. To ensure naturalness of 3N LECs, we further impose constraints on their values by requiring the expectation values of $V_{3\mathrm{N}}^{(c_D)}$ and $V_{3\mathrm{N}}^{(c_E)}$ no more than 30\% of the 2N energy. The fit result and  prediction of binding energy per nucleon are presented in Figure~\ref{fig:E_fit}. For comparison, we also show the experimental data and the results based on the 3N LECs determined in Ref.~\cite{Elhatisari:2022}. In general, our results deviate more  from the experiment compared to Ref.~\cite{Elhatisari:2022}. This is understandable, as the inclusion of $\mu$  and $\Delta E$ poses stringent constraints on the spin-orbit and tensor components of the 3N forces, which improves the prediction of phenomena sensitive to shell structures, such as $^{14}$C decay studied here. The trade-off, however, is a potential deterioration in the description of 3N components key to bulk properties such as the energy and collective behavior such as  $\alpha$-clustering. This likely accounts for the significant shifts observed in the predicted binding energies of $^{12}$C and $^{16}$O with pronounced $\alpha$-clustering structures. Nevertheless, the energies based on the optimized 3N force are within $2$-$3\sigma$s from the experiment for most nuclei, showing an overall acceptable agreement.

\begin{figure*}
\centering
\includegraphics[height=3.2in]{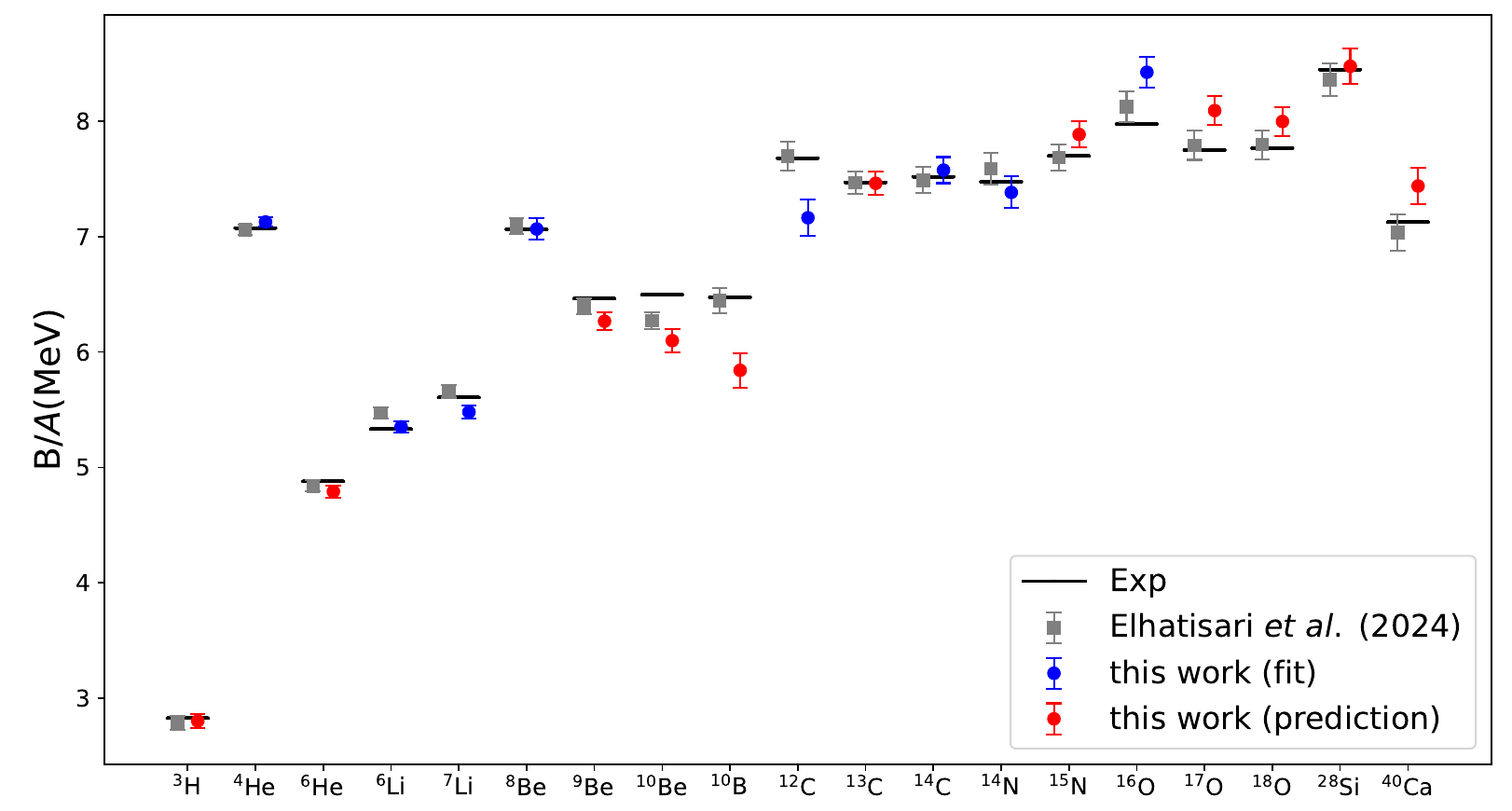}
\caption{Binding energy per nucleon for selected nuclei. The blue circles represent the nuclei used for the fit in this work and the red circles are predictions. The grays squares represent the result of Ref.~\cite{Elhatisari:2022}. The experimental values are given by the black lines.}
\label{fig:E_fit}
\end{figure*}

\begin{figure*}
\centering
\includegraphics[height=2.0in]{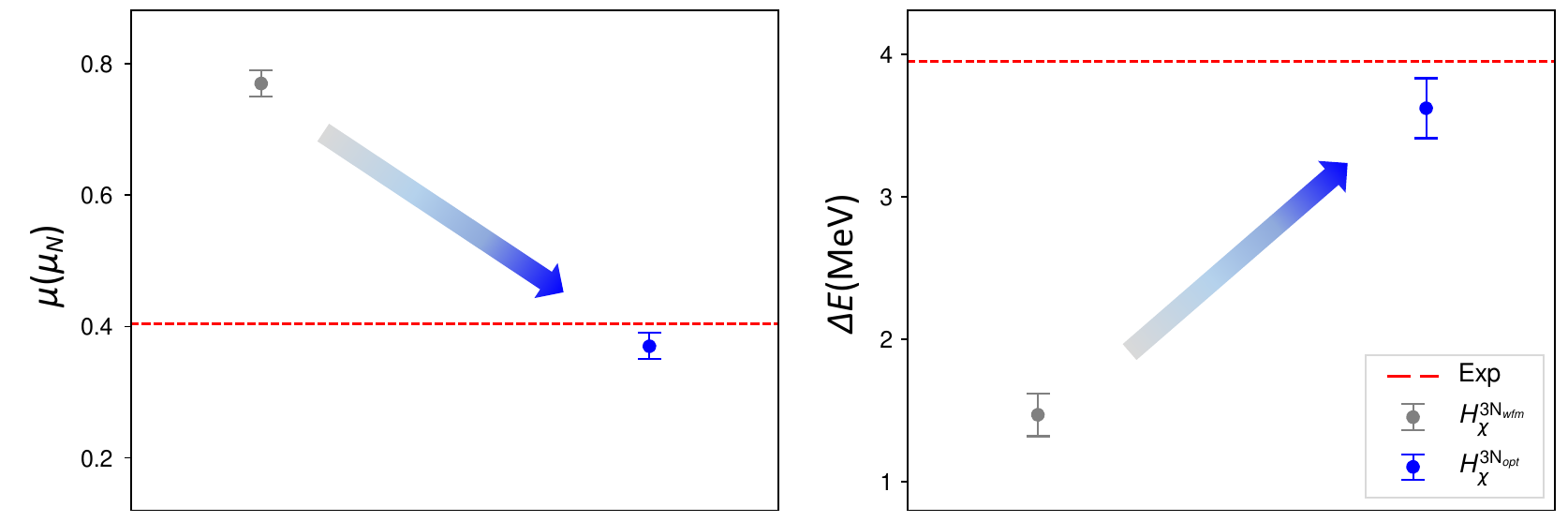}
\caption{(\textbf{Left panel}) The magnetic dipole moment $\mu$ of $^{14}$N calculated from $H_{\chi}^{3\mathrm{N}_{\mathrm{wfm}}}$ and $H_{\chi}^{3\mathrm{N}_{\mathrm{opt}}}$, denoted by the gray and blue points, respectively. (\textbf{Right panel}) The energy gap $\Delta E$ between $^{14}$N$(1^+_1)$ and $^{14}$N$(1^+_2)$ calculated from $H_{\chi}^{3\mathrm{N}_{\mathrm{wfm}}}$ and $H_{\chi}^{3\mathrm{N}_{\mathrm{opt}}}$, denoted by the gray and blue points, respectively. The red dashed lines represent the experimental values.}
\label{fig:mu_DeltaE_fit}
\end{figure*}

In Figure~\ref{fig:mu_DeltaE_fit}, we show how the optimization of 3N forces improves the prediction of $\mu$ and $\Delta E$. Without retuning 3N forces, the lattice result is far off the experimental value, showing the deficiency of the previous Hamiltonian $H_{\chi}^{3\mathrm{N}_{\mathrm{opt}}}$ in describing the shell structure of $^{14}$N. After the refit, the lattice result and the experiment almost match each other, reflecting the improvement of the $^{14}$N valence structures. Therefore, the incorporation of $\mu$ and $\Delta E$ into the fit is essential for correctly predicting $^{14}$C  $\beta$ decay.

\subsection{S3. GT transition operators and  matrix element}

\label{sec:S5}

The GT transition operators $O_{\mathrm{GT}}^{\mathrm{LO}}$ and $O_{\mathrm{GT}}^{\mathrm{>LO}}$ originate from the nuclear axial current $\boldsymbol{A}$. For the construction and explicit expressions of nuclear axial currents in the framework of $\chi$EFT, see Refs.~\cite{Krebs:2017,Girlanda:2016, Baroni:2017, Krebs:2020, Krebs2:2020}. In this work, we employ nuclear axial currents up to N$^3$LO, which consist of a LO term $\boldsymbol{A}^{\mathrm{LO}}$, a N$^2$LO relativistic correction term $\boldsymbol{A}^{\mathrm{N}^2\mathrm{LO}}$, as well as a OPE term $\boldsymbol{A}^{\mathrm{N}^3\mathrm{LO}}(\mathrm{OPE})$ and a contact term $\boldsymbol{A}^{\mathrm{N}^3\mathrm{LO}}(\mathrm{CT})$ at N$^3$LO, following the power counting rule used in our previous work~\cite{Teng:2025}. Their diagrammatic illustrations are represented in Figure~\ref{fig:current}. The relationship between $O_{\mathrm{GT}}$ and $\boldsymbol{A}$ follows the same convention  adopted in Ref.~\cite{Teng:2025}: 
\begin{equation} 
\begin{aligned}
&O_{\mathrm{GT},\lambda}^{\mathrm{LO}} = A_{\lambda}^{\mathrm{LO}},\\
& O_{\mathrm{GT},\lambda}^{>\mathrm{LO}} = A_{\lambda}^{\mathrm{N}^2\mathrm{LO}}+A_{\lambda}^{\mathrm{N}^3\mathrm{LO}}(\mathrm{OPE})+A_{\lambda}^{\mathrm{N}^3\mathrm{LO}}(\mathrm{CT}).
\end{aligned}
\end{equation}
For the definition of the GT matrix element $M_{\mathrm{GT}}$, we also follow Ref.~\cite{Teng:2025},
\begin{equation}
\label{eq:MGT_definition}
    M_{\mathrm{GT}} = \frac{\sqrt{2J_f+1}}{{g}_A}\left.\frac{\langle f|O_{\mathrm{GT},\lambda}|i\rangle}{\langle J_i M_i; 1\lambda|J_f M_f\rangle }\right|_{M_f = M_i+\lambda},
\end{equation}
where $|i\rangle$ and $|f\rangle$ denote the initial and final states, in order. $(J_{i}, M_i)$ and $(J_f, M_f)$ denote their respective  angular momenta and $\langle J_i M_i; 1\lambda|J_f M_f\rangle $ is the Clebsch-Gordan coefficient.  The constraint $M_f = M_i+\lambda$ is imposed for angular momentum conservation. The transition matrix element $\langle f|O_{\mathrm{GT},\lambda}|i\rangle$ in Eq.~(\ref{eq:MGT_definition}) is  extracted from the following ratio,
\begin{equation}
    \langle f|O_{\mathrm{GT},\lambda}|i\rangle=\lim_{\tau\rightarrow\infty}\frac{\boldsymbol{v}^{\dagger}_{\mathrm{N}} M_{\mathrm{GT}, \lambda}(\tau) \boldsymbol{v}_{\mathrm{C}}}{\sqrt{[\boldsymbol{v}^{\dagger}_{\mathrm{C}}N_{\mathrm{C}}(\tau)\boldsymbol{v}_{\mathrm{C}}] [\boldsymbol{v}^{\dagger}_{\mathrm{N}}N_{\mathrm{N}}(\tau)\boldsymbol{v}_{\mathrm{N}}]}}
\end{equation}
where the vectors $\boldsymbol{v}_{\mathrm{C}}$ and $\boldsymbol{v}_{\mathrm{N}}$ are the ground-state solutions of the generalized eigenvalue equation~(\ref{eq: GEVP}).  $N_{\mathrm{C}/\mathrm{N}}(\tau)$ and  $M_{\mathrm{GT}, \lambda}(\tau)$ are matrix abbreviations of the correlation functions defined in Eq.~(\ref{eq:N_ij}) and~(\ref{eq: MGT_ij}), respectively. 

\begin{figure*}
\centering
\includegraphics[width=0.76\textwidth]{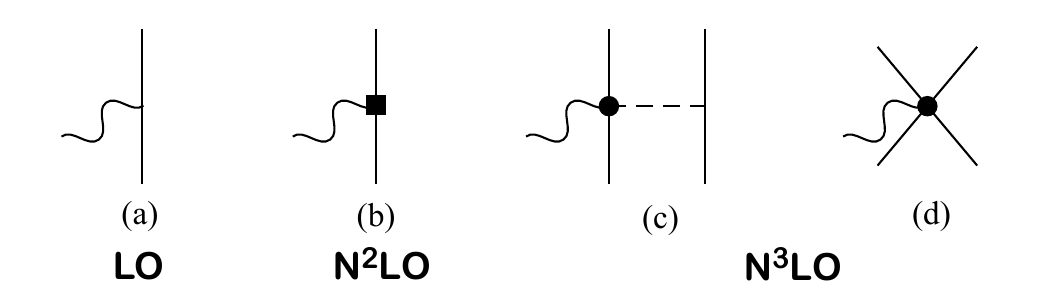}
\caption{Diagrammatic illustration of chiral axial currents up to N$^3$LO which contribute to the GT transition operator used in this work. Nucleons, 
pions and the external field are denoted by the solid,
dashed, and wavy lines, respectively. Panel (a) denotes the LO current while panel (b) gives the relativistic correction at N$^2$LO. Panels (c) and (d) display the OPE current and the contact current at N$^3$LO, respectively.} \label{fig:current}
\end{figure*}

\subsection{S4. Shell-model trial states}
\label{sec:S3}
We employ shell-model trial states for the calculation of the $^{14}$C and the $^{14}$N ground state.  For each nucleus, we consider different configurations  of valence nucleons, and each configuration $i$ corresponds to a trial state $|\Psi_{T,\mathrm{C}/\mathrm{N}}^i\rangle$. To ensure that the trial state carries the desired spin $J$ and parity $\pi$, $|\Psi_{T,\mathrm{C}/\mathrm{N}}^i\rangle$ is constructed as a linear combinations of single Slater determinants. For more details, we refer the reader to Ref.~\cite{Wang:2025_MR}.  In this work, we use 5 different configurations of $^{14}$C$(0^+)$ and 4 different configurations of $^{14}$N$(1^+)$ in total to construct the trial states, whose diagrammatic illustrations are given in Figure~\ref{fig:C14_trial_state}
and Figure~\ref{fig:N14_trial_state}, respectively.

\begin{figure*}
\centering
\includegraphics[width=0.76\textwidth]{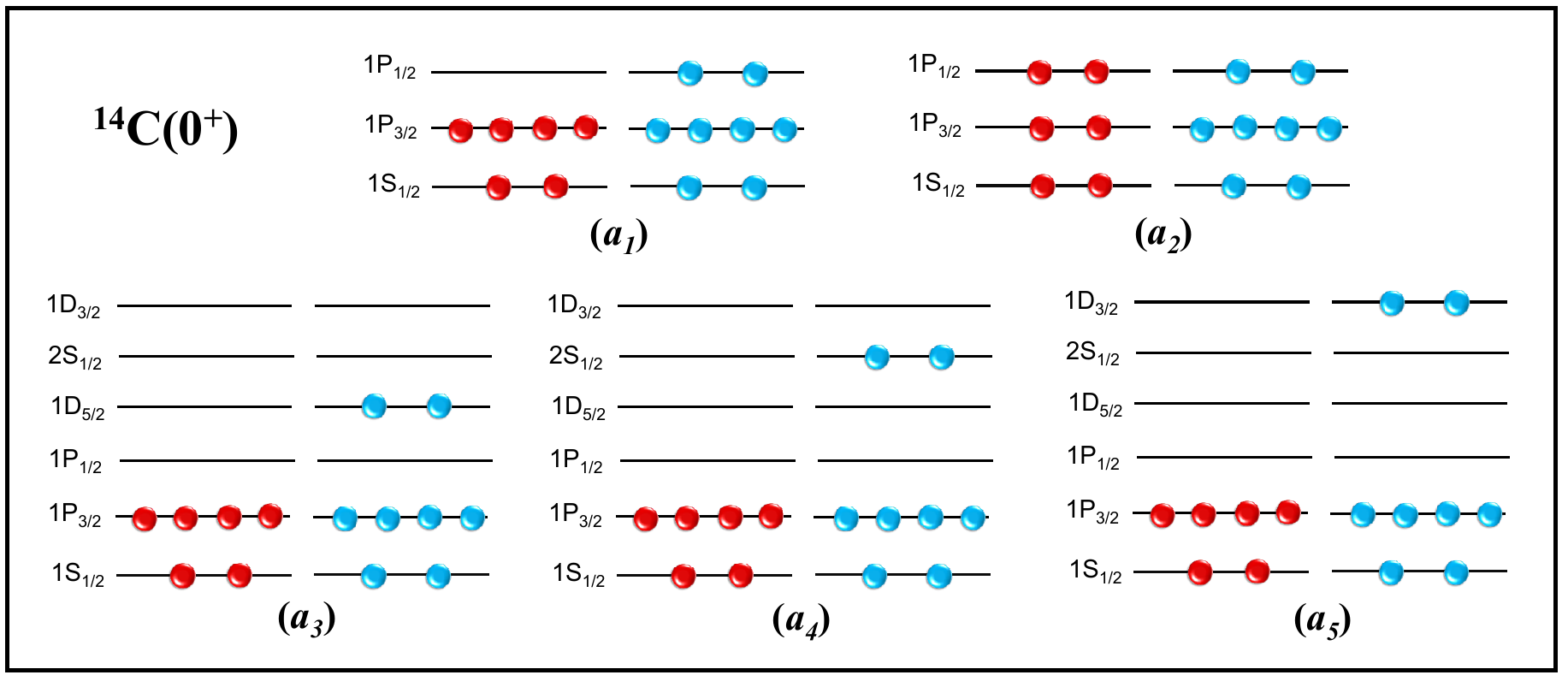}
\caption{Diagrammatic illustration for the different trial states of $^{14}$C$(0^+)$ used in this work. Subfigures $(a_1)$-$(a_5)$ represent different shell model configurations composed of linear combinations of single Slater determinants, ensuring that each configuration carries the correct quantum numbers $J^\pi=0^+$. For each configuration, the labels on the left denote single particle shell-model orbits, while the red and blue circles represent protons and neutrons, respectively. } \label{fig:C14_trial_state}
\end{figure*}

\begin{figure*}
\centering
\includegraphics[width=0.76\textwidth]{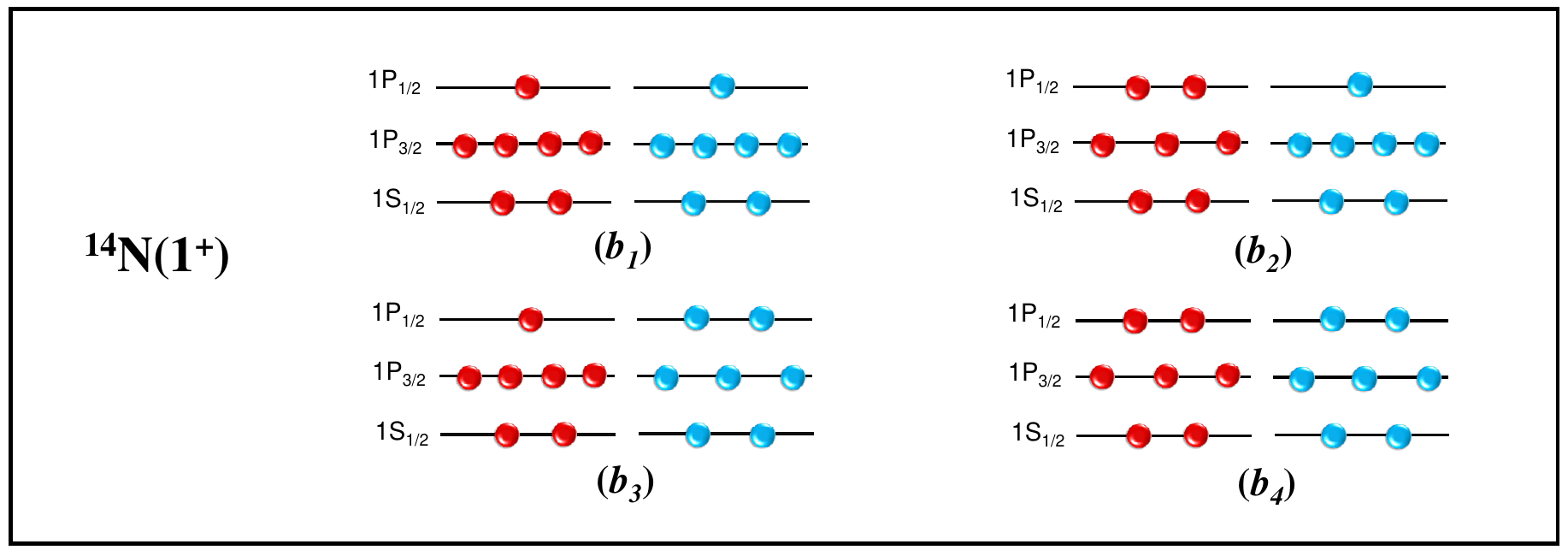}
\caption{Diagrammatic illustration for the different trial states of $^{14}$N$(1^+)$ used in this work. Subfigures $(b_1)$-$(b_4)$ represent different shell model configurations composed of linear combinations of single Slater determinants, ensuring that each configuration carries the correct quantum numbers $J^\pi=1^+$. For each configuration, the labels on the left denote single particle shell-model orbits, while the red and blue circles represent protons and neutrons, respectively.} \label{fig:N14_trial_state}
\end{figure*}

\subsection{S5. Lattice Monte Carlo calculation of correlation functions}

\label{sec:S4}

The multi-channel variational method requires the input of correlation functions between different trial states, as is discussed in the main text. In the following, we first discuss how to calculate these correlation functions through lattice quantum Monte Carlo. Then, we detail the important improvement we made to optimize the calculation.

In realistic NLEFT calculations, the imaginary-time projection operator $e^{-H_S\tau}$  is decomposed into the multiplication of $N_t=\tau/a_t$ transfer matrices $M=:e^{-H_S a_t}:$, with $a_t$ the temporal lattice spacing. Applying auxiliary transformation to the transfer matrices, we obtain,
\begin{equation}
    e^{-H_S\tau}=\int \mathcal{D}\xi \ e^{-\sum_{n_t=1}^{N_t}\xi^2(n_t)/2}\prod_{n_t=1}^{N_t}\mathcal{M}[\xi(n_t)],
\end{equation}
where $\xi(n_t)$ denotes the collection of auxiliary fields on the $n_t$th time slice, $\mathcal{D}\xi$ denotes the integral measure and $\mathcal{M}[\xi(n_t)]$ is the transformed transfer matrix. Operating $\mathcal{M}[\xi]$ successively on  the trial states $|\Psi_{T,\mathrm{C}/\mathrm{N}}^i\rangle$, we define the following ket and bra,
\begin{equation}
\begin{aligned}
    &|\Phi^{i,R}_{\tau, \mathrm{C}/\mathrm{N}}(\xi)\rangle = \left(\prod_{N_t=1}^{L_t/2}\mathcal{M}[\xi(n_t)]\right)|\Psi^i_{T,\mathrm{C}/\mathrm{N}}\rangle, \\
   &\langle\Phi^{i,L}_{\tau,\mathrm{C}/\mathrm{N}}(\xi)|= \langle\Psi^i_{T,\mathrm{C}/\mathrm{N}}|\left(\prod_{n_t=N_t/2+1}^{L_t}\mathcal{M}[\xi(n_t)]\right),
\end{aligned}
\end{equation}
based on which the correlation functions in Eq.~(\ref{eq:N_ij}),~(\ref{eq:H_ij})  and (\ref{eq: MGT_ij}) can be expressed as,
\begin{equation}
    \begin{aligned}
        &N^{ij}_{\mathrm{C}/\mathrm{N}}(\tau) = \int \mathcal{D}\xi\ e^{-\xi^2/2} \langle\Phi^{i,L}_{\tau,\mathrm{C}/\mathrm{N}}(\xi)|\Phi^{j,R}_{\tau, \mathrm{C}/\mathrm{N}}(\xi)\rangle,\\
        &H^{ij}_{\chi, \mathrm{C}/\mathrm{N}}(\tau) = \int \mathcal{D}\xi\ e^{-\xi^2/2} \langle\Phi^{i,L}_{\tau,\mathrm{C}/\mathrm{N}}(\xi)|H_\chi|\Phi^{j,R}_{\tau, \mathrm{C}/\mathrm{N}}(\xi)\rangle,\\
        &M_{\mathrm{GT},\lambda}^{ij}(\tau) = \int  \mathcal{D}\xi\ e^{-\xi^2/2} \langle\Phi^{i,L}_{\tau, \mathrm{N}}(\xi)|O_{\mathrm{GT},\lambda}|\Phi^{j,R}_{\tau, \mathrm{C}}(\xi)\rangle.
    \end{aligned}
\end{equation}
To compute the integrals over $\xi$, a proper probability distribution function $P(\xi)$ is chosen to generate  configurations of auxiliary fields. The high-dimensional integral can be replaced by summation over the configurations, which can be done numerically~\cite{Lahde:2019, Wang:2025_MR}.

When calculating the multi-channel correlation functions, we found two major obstacles that severely hinder the  computation: First, an efficient sampling algorithm is required to generate auxiliary field configurations, otherwise the computational cost for the multi-channel calculation would be expensive as is discussed in Ref.~\cite{Wang:2025_MR}. Second,  due to the complexity of the higher-order nuclear forces, the calculation of matrix elements of the high-fidelity Hamiltonian $H_\chi$ between single Slater determinants is time consuming. Though it is acceptable for one-channel case, in our multi-channel calculation, the different trial states of each nucleus contain  around 10 Slater determinants, hence there are nearly 200 matrix elements of $H_\chi$ to calculate, making the computation extremely challenging.

To address the first issue, we employ  the algorithm developed in Ref.~\cite{Wang:2025_MR}. The  weighting function for generating auxiliary field configurations is chosen as
\begin{equation}
\begin{aligned}
     P(\xi)&\propto e^{-\xi^2/2}\Big{[}\sum_{i}a_i|\langle \Psi^i_{L,\mathrm{C}}(\xi)|\Psi^i_{R,\mathrm{C}}(\xi)\rangle|+\sum_{i}b_i|\langle \Psi^i_{L,\mathrm{N}}(\xi)|\Psi^i_{R,\mathrm{N}}(\xi)\rangle|\Big{]},
\end{aligned}
\end{equation}
where $a_i$ and $b_i$ are weight factors tuned to control statistical fluctuations. The auxiliary fields are then sampled using the shuttle algorithm~\cite{Lu:2018}. We exploit the common orbits shared by different trial states to design the code, greatly accelerating the propagation of states in the shuttle algorithm~\cite{Wang:2025_MR}.

Concerning the second issue, we optimized the numerical function for calculating the matrix element of $H_\chi$, based on the original code developed by the NLEFT collaboration. Detailed improvements include replacing loop iterations with array operations, employing fast Fourier transformation to compute convolutions and optimizing data structure to avoid repeated calculations. After the optimization, the calculation of $H_\chi$ on CPU is accelerated by $50$-$60$ times,  which greatly reduces the computational time  from several years to a few weeks.

\subsection{S6. Fit of the GT matrix element}

\label{sec:S6}

In Figure~\ref{fig:GT_fit}, we show the imaginary-time evolution and fitting result of $M_{\mathrm{GT}}$  for the 4 different combinations of interactions and transition operators, corresponding to the 4 data points shown in Figure~\ref{fig:GT_vs_interaction}. A constant fit is employed for the data points belonging to the last 5 time slices. 

\begin{figure*}
\centering
\includegraphics[width=0.76\textwidth]{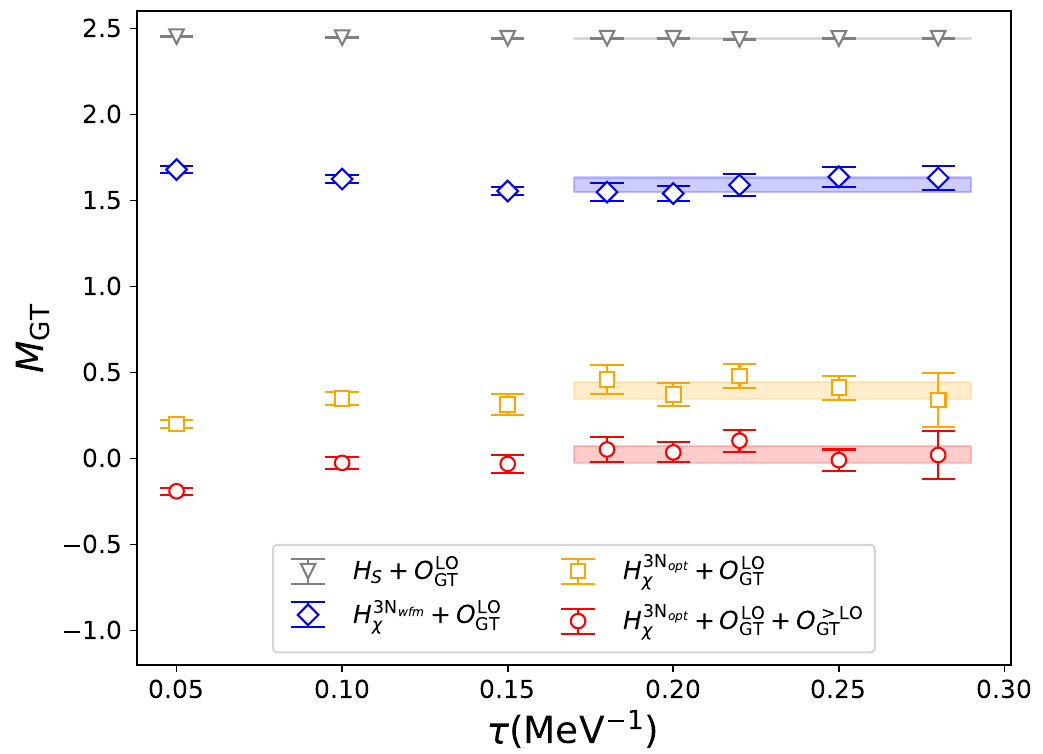}
\caption{The imaginary-time evolution and fit result of $M_{\mathrm{GT}}$  for the different combinations of interactions and transition operators. The gray triangle, blue diamond and yellow square represent the data calculated with the LO GT operator $O_{\mathrm{GT}}^\mathrm{LO}$, as well as Hamiltonians $H_S$, $H_\chi^{3\mathrm{N}_{\mathrm{wfm}}}$ and $H_\chi^{3\mathrm{N}_{\mathrm{opt}}}$ respectively.  The red circle additionally includes the correction from the higher-order GT operator $O^{>\mathrm{LO}}_{\mathrm{GT}}$. The error bar denotes the statistical uncertainty. The bands of different colors represent the fit ranges and results.}
\label{fig:GT_fit}
\end{figure*}

\subsection{S7. Finite volume analysis}

\label{sec:S7}

To analyze the finite volume dependence of $M_{\mathrm{GT}}$, we performed calculations for five different boxes with lengths  $L$ ranging from $7a$ to $11a$, with the lattice spacing $a=1.32$ fm. The fit results of $M_{\mathrm{GT}}$ are shown in Figure~\ref{fig:GT_FV}. A plateau appears for $L\ge 8a$, indicating that finite volume artifacts are already negligible compared to the statistical errors, and the result of $L=11a$ is used throughout the main text.

\begin{figure*}
\centering
\includegraphics[width=0.76\textwidth]{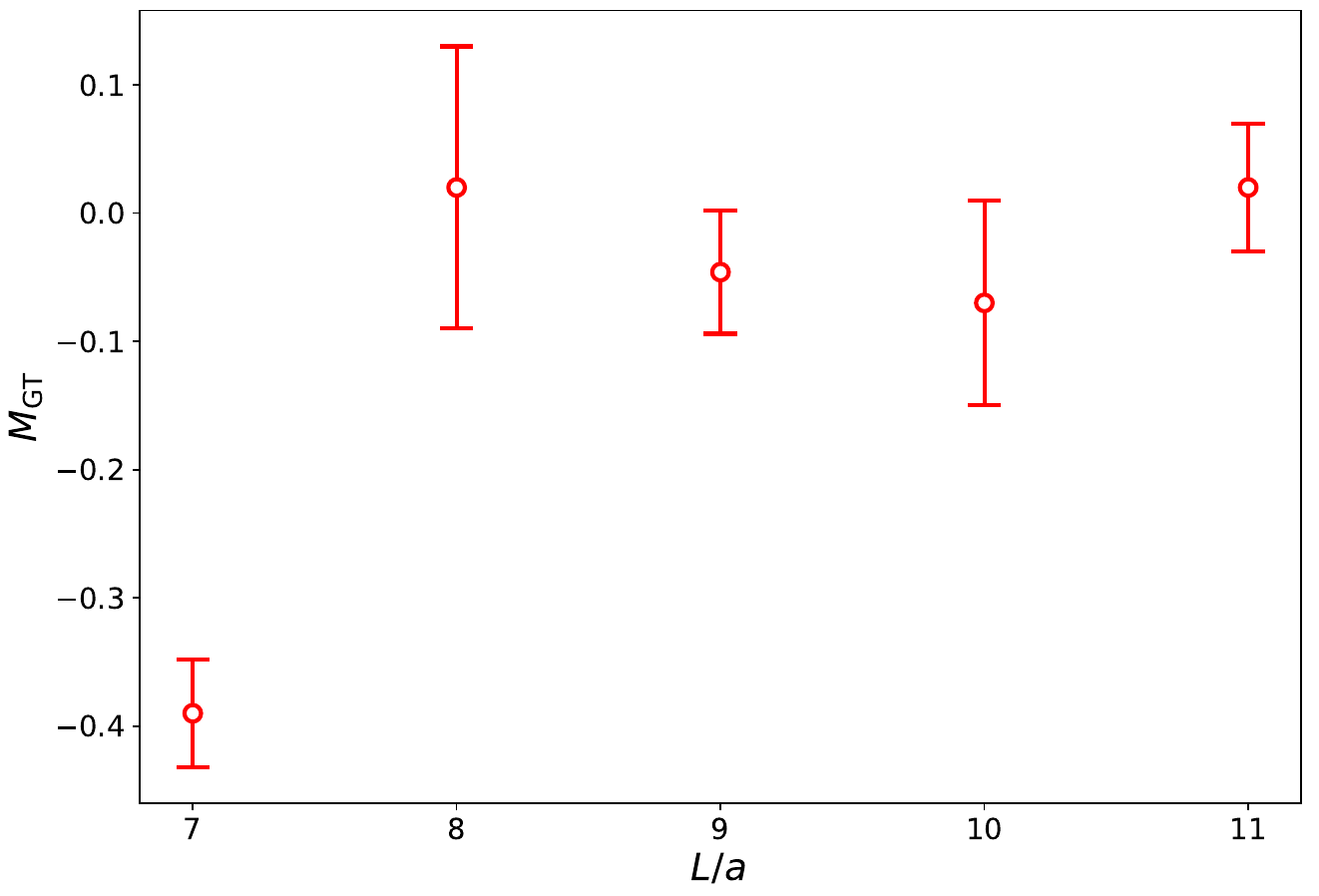}
\caption{The fit results of $M_{\mathrm{GT}}$ for boxes with different lengths $L$. $a=1.32$ fm is the lattice spacing. The error bar represents the statistical uncertainty.}
\label{fig:GT_FV}
\end{figure*}

\subsection{S8. The  projection operator $P_\Gamma$}

\label{sec:S9}

The projection operator $P_\Gamma$ in Eq.~(\ref{eq:irrep_proj}) takes the following form,
\begin{equation}
    \begin{aligned}
        P_\Gamma = \frac{d_\Gamma}{N}\sum_{R\in O} \chi_\Gamma(R) P(R),
    \end{aligned}
\end{equation}
where $N=24$ is the total number of elements $R$ in the octahedral group $O$, $d_\Gamma$ is the dimension of the irrep $\Gamma$, $\chi_\Gamma(R)$ is the  character of $R$ for $\Gamma$ and $P(R)$ is the corresponding rotational  operator. In Table~\ref{tab:chraracter}, we provide the values of $d_\Gamma$ and $\chi_\Gamma(R)$.

\begin{table}[htbp]
\begin{center}
\setlength{\tabcolsep}{2.5mm}
		\resizebox{250pt}{!}{\begin{tabular}{c| c |c |c |c |c }
			\hline\hline
			       & $I$ & 8$C_3$& 3$C_2$ & 6$C'_2$& 6$C_4$   \\
                 \hline
                $A_1(d_{A_1}=1)$ & 1&1&1&1&1\\ 
		\hline
  $A_2(d_{A_2}=1)$ & 1&1&1&-1&-1\\
  \hline
  $E(d_E=2)$ & 2&-1&2&0&0\\
  \hline
  $T_1(d_{T_1}=3)$ & 3&0&-1&-1&1\\ 
  \hline
  $T_2(d_{T_2}=3)$ & 3&0&-1&1&-1\\ 
  \hline\hline
		\end{tabular}}
  \end{center}
		\caption{Table of characters $\chi_\Gamma$ of the octahedral group O, as well as the dimension $d_\Gamma$ of of each irrep $\Gamma$. The first column shows the five different irreps of $O$, and their respective dimensions $d_\Gamma$. The last five volumes show the characters $\chi_\Gamma(R)$. The first row contains all the 24 elements of $O$, including the identity $I$, eight rotations about cube body diagonals ($8C_3$), nine rotations
around the $x,y,z$ axes ($3C_2$ and $6C_4$), and six rotations about axes parallel to face diagonals
($6C'_2$).   }
	\label{tab:chraracter}
\end{table}

\subsection{S9. Duality between the NLEFT result and the shell model scenario}

\label{sec:S12}

From the perspective of the nuclear shell model, the ground states of $^{14}$C and $^{14}$N can be viewed as an $^{16}$O-core with $J^P=0^+$ plus two holes in the $p$-shell. Due to the constraint of spin, isospin and parity ($J^P=0^+, T=1$ for $^{14}$C and $J^P=1^+, T=0$ for $^{14}$N), they can be expressed as linear combinations of the partial wave eigenstate $|^{2S+1}L_J\rangle$~\cite{Jancovici:1954, Talmi:2022},
\begin{equation}  \label{eq:SM_eigenstates} |^{14}\mathrm{C}\rangle = x|^{1}S_0\rangle +y|^3P_0\rangle,\quad |^{14}\mathrm{N}\rangle = \alpha |^3 S_1\rangle+\beta|^3 D_1\rangle +\gamma|^1 P_1 \rangle.
\end{equation}
The magnitude of $M_{\mathrm{GT}}$ depends on the  values of the coefficients $(x,y)$ and $(\alpha,\beta,\gamma)$, which can be determined by diagonalizing the shell-model Hamiltonian matrix. Note that for the LO GT transition operator $O^{\mathrm{LO}}_{\mathrm{GT}}$ with no dependence on the orbital degree of freedom, its matrix element between partial wave eigenstates with different values of $L$ is strictly forbidden.  

\begin{figure*}
\centering
\includegraphics[width=0.594\textwidth]{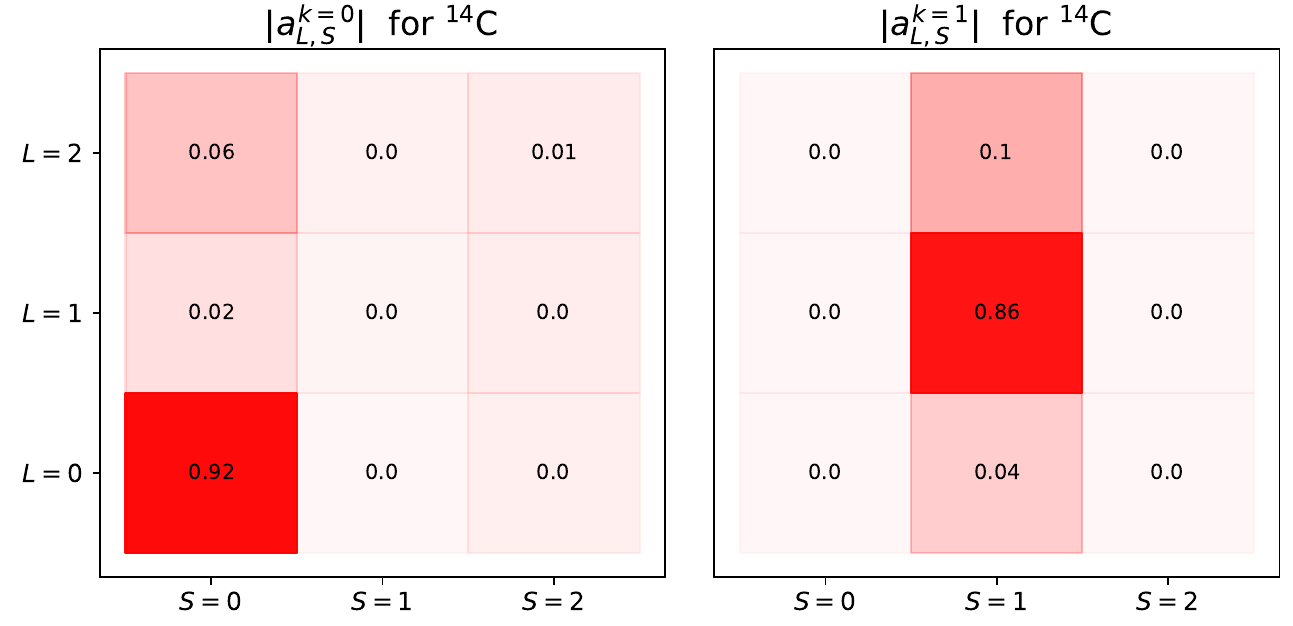}
\caption{The distribution of $|a_{L,S}^{k=0, 1}|$ for $^{14}$C. The number in each square denotes the corresponding value of $|a_{L,S}^{k}|$.}
\label{fig:aLS_C14}
\end{figure*}

\begin{figure*}
\centering
\includegraphics[width=0.888\textwidth]{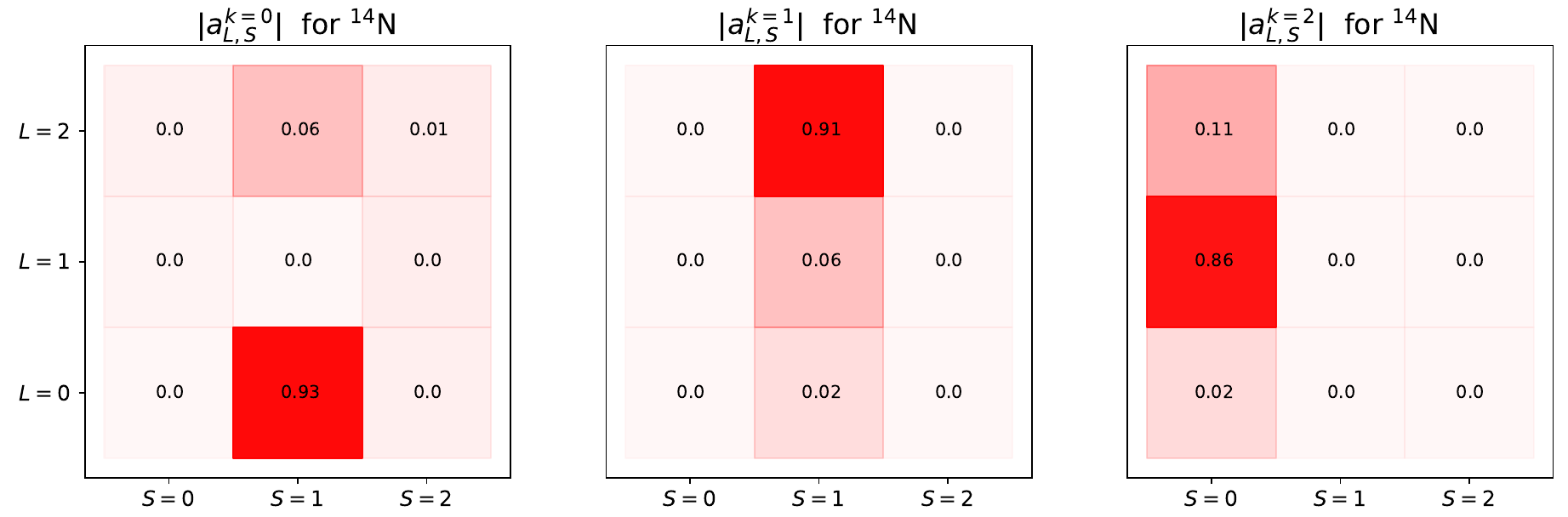}
\caption{The distribution of $|a_{L,S}^{k=0,1,2}|$ for $^{14}$N. The number in each square denotes the corresponding value of $|a_{L,S}^{k}|$. }
\label{fig:aLS_N14}
\end{figure*}

In our NLEFT calculation, the multi-channel variational method allows  to express the ground state $|\Phi^0_{\mathrm{C}/\mathrm{N}}\rangle$ of $^{14}$C and  $^{14}$N in terms of the low-lying eigenstates $|\Psi^k_{\mathrm{C}/\mathrm{N}}\rangle$ of  the LO chiral Hamiltonian $H_S$ ($k$ denotes the $k$th excited state), 
\begin{equation}
   \label{eq:NLEFT_eigenstates} |\Phi^0_{\mathrm{C}}\rangle = \tilde{x}|\Psi^0_{\mathrm{C}}\rangle+\tilde{y}|\Psi^1_{\mathrm{C}}\rangle+\cdots,\quad |\Phi^0_{\mathrm{N}}\rangle = \tilde{\alpha}|\Psi^0_{\mathrm{N}}\rangle+\tilde{\beta}|\Psi^1_{\mathrm{N}}\rangle+\tilde{\gamma}|\Psi^2_{\mathrm{N}}\rangle+\cdots,
\end{equation}
where $\cdots$ denotes higher excited states whose contributions are found to be small. Since $H_S$ is dominated by the strong SU(4) symmetric term $V_{\mathrm{SU}(4)}$ in Eq.~(\ref{eq:H_S_expression}), while the remaining SU(4)-breaking terms are weak,  the total orbital angular momentum $L$ and spin $S$ can be treated as approximate good quantum numbers for $H_S$,  and one may anticipate that there would be a duality between these eigenstates of $H_S$ and the shell model state $|^{2S+1}L_J\rangle$. To verify this assumption,  we calculate the proportion of different $L$- and $S$-components in $|\Psi^k_{\mathrm{C}/\mathrm{N}}\rangle$ by calculating 
\begin{equation}
    a^k_{L,S} = \langle \Psi_{\mathrm{C/\mathrm{N}}}^k|P_S P_L|\Psi_{\mathrm{C/\mathrm{N}}}^k\rangle.
\end{equation}
In the above, $P_L$ is the lattice orbital-angular-momentum projection operator constructed from $P_\Gamma$ according to Table~\ref{tab:irrep_dec},
\begin{equation*}
    P_{L=0} = P_{A_1},\quad P_{L=1} = P_{T_1},\quad P_{L=2} = P_{E}+P_{T_2},
\end{equation*}
and $P_S$ is the projection operator of the total spin,
\begin{equation*}
    P_S = \frac{2S+1}{8\pi^2}\sum_{S_z}\int d\Omega \ \left[D^{S}_{S_z S_z}(\Omega)\right]^*R(\Omega),
\end{equation*}
with $R(\Omega)$ the rotational operator acting on the spin degree of freedom and $D^{S}_{S_z S_z}(\Omega)$ the Wigner-D function.  In Figure~\ref{fig:aLS_C14} and~\ref{fig:aLS_N14}, we show the distribution of the absolute value of $a_{L,S}$ for $|\Psi^{k=0,1}_{\mathrm{C}}\rangle$ and $|\Psi^{k=0,1, 2}_{\mathrm{N}}\rangle$. Clearly, each of these eigenstates of $H_S$ is dominated by a specific value of $L$ and $S$, and there is an interesting mapping between these eigenstates  and the shell model partial wave states in Eq.~(\ref{eq:SM_eigenstates}),
\begin{equation}
\label{eq:NLEFT_SM_mapping}
\begin{aligned}    &|\Psi^0_{\mathrm{C}}\rangle\mapsto |^1S_0\rangle,\quad |\Psi^1_{\mathrm{C}}\rangle\mapsto |^3P_0\rangle, \\
& |\Psi^0_{\mathrm{N}}\rangle\mapsto |^3S_1\rangle,\quad |\Psi^1_{\mathrm{N}}\rangle\mapsto |^3D_1\rangle ,\quad |\Psi^2_{\mathrm{N}}\rangle\mapsto |^1P_1\rangle.
\end{aligned}
\end{equation}
The mapping above validates the shell-model ansatz on the ground state structures of $^{14}$C and $^{14}$N. It also offers a convenient tool for understanding the mechanism of $^{14}$C $\beta$ decay discussed in the main text, as illustrated below.

\begin{figure*}
\centering
\includegraphics[width=0.756\textwidth]{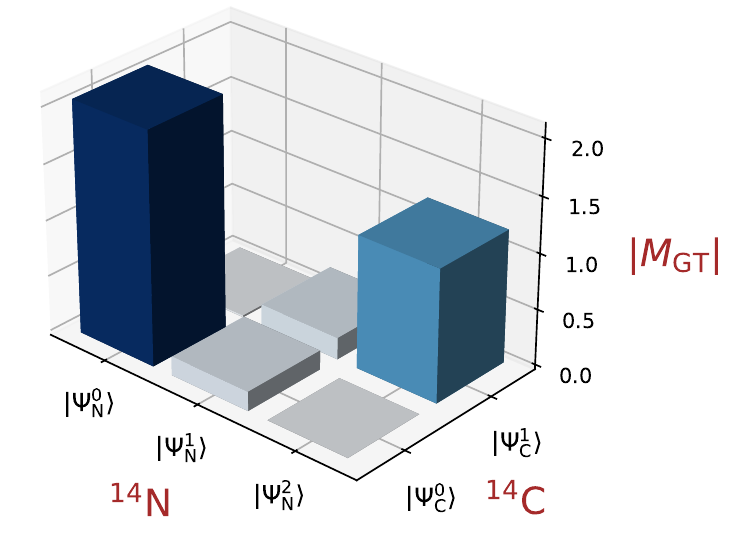}
\caption{The absolute value distribution of the Gamow-Teller matrix element $M_{\mathrm{GT}}$ between the initial state $|\Psi^{k}_{\mathrm{C}}\rangle$ and the final state $|\Psi^{k'}_{\mathrm{N}}\rangle$.  }
\label{fig:GT_between_HS_eigenstates}
\end{figure*}

In Figure~\ref{fig:GT_between_HS_eigenstates}, we show the distribution of the GT matrix element for $O^{\mathrm{LO}}_{\mathrm{GT}}+O^{\mathrm{>LO}}_{\mathrm{GT}}$ between $|\Psi^k_{\mathrm{C}}\rangle$ and $|\Psi^{k'}_{\mathrm{N}}\rangle$. It can be seen that the transition is enhanced in $|\Psi^0_{\mathrm{C}}\rangle\rightarrow |\Psi^0_{\mathrm{N}}\rangle$ and $|\Psi^2_{\mathrm{C}}\rangle\rightarrow |\Psi^1_{\mathrm{N}}\rangle$ channels but greatly suppressed in the other four channels. According to the mapping in Eq.~(\ref{eq:NLEFT_SM_mapping}), this  observation is consistent with the shell model interpretation that only the $S$- and $P$-wave transitions are allowed. The reasons why the transitions of the other four channels are not strictly forbidden are twofold: 1. $H_S$ is not purely SU(4) symmetric so $|\Psi^{k}_{\mathrm{C}/\mathrm{N}}\rangle$ is not a rigorous eigenstate of $L$ and $S$; 2.the matrix element shown in Fig.~\ref{fig:GT_between_HS_eigenstates} includes corrections from higher-order GT transition operators.

The full GT matrix element between $|\Phi^0_{\mathrm{C}}\rangle$ and $|\Phi^0_{\mathrm{N}}\rangle$ is determined by the coefficients $(\tilde{x},\tilde{y})$ and $(\tilde{\alpha},\tilde{\beta},\tilde{\gamma})$ (contributions from even higher excited states of $H_S$ are negligible), which  can be solved from the Hamiltonian matrix of $H_\chi$ in the basis of $|\Phi^0_{\mathrm{C}/\mathrm{N}}\rangle$. Notably, we find the values of the $^{14}$N coefficients $(\tilde{\alpha},\tilde{\beta},\tilde{\gamma})$ sensitive to the interaction employed. Specifically, we define the subtracted Hamiltonian matrix $\overline{H_\chi}$ for $^{14}$N,
\begin{equation}   \left(\overline{H_\chi}\right)_{k,k'} = \langle\Phi^k_{\mathrm{N}}|H_\chi|\Phi^{k'}_{\mathrm{N}}\rangle - \delta_{k,k'}\langle\Phi^0_{\mathrm{N}}|H_\chi|\Phi^0_{\mathrm{N}}\rangle.
\end{equation}
Note that the subtraction of the second part does not influence the solution of $(\tilde{\alpha},\tilde{\beta},\tilde{\gamma})$, and helps to better visualize the energy gaps between different eigenstates of $H_S$ with respect to $H_\chi$.  We also define the subtracted potential matrix $\overline{V}$ for $^{14}$N, 
\begin{equation}  \left(\overline{V}\right)_{k,k'} = \langle\Phi^k_{\mathrm{N}}|V|\Phi^{k'}_{\mathrm{N}}\rangle - \delta_{k,k'}\langle\Phi^0_{\mathrm{N}}|V|\Phi^0_{\mathrm{N}}\rangle.
\end{equation}
For $H_\chi = H_{\chi,2\mathrm{N}}+V_{\mathrm{3\mathrm{N}}}$, the subtracted matrix of its two-body part is (statistical uncertainties do not influence the discussions and are omitted below) 
\begin{equation}
\overline{H_{\chi,2\mathrm{N}}} = \begin{bmatrix}
    0&   -1.32& -4.46\\
    -1.32& 1.98 &
-4.44\\
-4.46& -4.44& 16.31\end{bmatrix}(\mathrm{MeV}),
\end{equation}
which means that its expectation energy with respect to the $^3S_1$-configuration is lower than the  $^1P_1$- and $^3D_1$-configurations (see. Eq.~(\ref{eq:NLEFT_SM_mapping})). Consequently, its solution $(\tilde{\alpha},\tilde{\beta},\tilde{\gamma}) = (0.79, 0.53, 0.31)$ is dominated by the $S$ wave. The matrix form of the old 3N force $V_{3\mathrm{N}}^{\mathrm{wfm}}$ is
\begin{equation}
\label{eq:mat_3Nold}
\overline{V_{3\mathrm{N}}^{\mathrm{wfm}}} = \begin{bmatrix}
    0&   1.61& 0.09\\
    1.61& -1.68 &
0.14\\
0.09& 0.14& 3.54\end{bmatrix}(\mathrm{MeV}),
\end{equation}
which bring the energy of $^3D_1$ configuration closer to $^3S_1$ configuration, but not enough. The resulting solution is $(\tilde{\alpha},\tilde{\beta},\tilde{\gamma}) = (0.76, 0.59, 0.28)$ and still populated by the $S$ wave, consistent with the second subplot in Figure~\ref{fig:irrep_vs_interaction}. This also explains why the second data point in Figure~\ref{fig:GT_vs_interaction} deviates from the experiment significantly. For the optimized 3N force $V_{\mathrm{3N}}^{\mathrm{opt}}$, its subtracted matrix is
\begin{equation}
 \label{eq:mat_3Nnew}  \overline{V_{3\mathrm{N}}^{\mathrm{opt}}}= \begin{bmatrix}
        0& 2.62&0.08\\
        2.62& -4.58& 0.14\\
        0.08& 0.14& 1.89
    \end{bmatrix}(\mathrm{MeV}),
\end{equation}
which gives the $^{3}D_1$-configuration a strong negative correction and reverses its order compared to the $^{3}S_1$-configuration. The coefficients for the optimized Hamiltonian $H_{\chi}^{3\mathrm{N}_{\mathrm{opt}}}$ are hence $(\tilde{\alpha},\tilde{\beta},\tilde{\gamma}) = (-0.19, 0.97,0.15)$, which are dominated by the $D$ wave and agree with the third subplot in Figure~\ref{fig:irrep_vs_interaction}. This also explains the drastic jump from the second to the third data point in Figure~\ref{fig:GT_vs_interaction}. We remark that although the absolute value of 3N energy is significantly smaller than the 2N energy,  the 3N force plays a critical role in controlling the order of energy levels between different shell model configurations, as can be seen in Eq.~(\ref{eq:mat_3Nold}) and (\ref{eq:mat_3Nnew}).  

Concerning the result of the GSA shown in Figure~\ref{fig:ST_distribution}, we also provide a qualitative explanation. The subtracted matrices of the two LO 2N potentials, $V_{^1S_0}$ and $V_{^3 S_1}$, are given by
\begin{equation}
    \overline{V_{^1S_0}} = \begin{bmatrix}
        0&0.02&0.00\\
        0.02&1.83&0.02\\
        0.00&0.02&3.05
    \end{bmatrix}(\mathrm{MeV}),\quad \overline{V_{^3S_1}} = \begin{bmatrix}
        0&0.20&0.02\\
        0.20&8.90&0.02\\
        0.02&0.02&15.38
    \end{bmatrix}(\mathrm{MeV}).
\end{equation}
Compared to $V_{^1S_0}$, $V_{^3S_1}$ provides a strong repulsion to the $^3D_1$-configuration. Therefore, in  the vicinity of $M_{\mathrm{GT}}=0$, the variation of $V_{^3S_1}$ would induce significant changes of the $S$- and $D$-wave components in the wave function of $^{14}$N, which explains its large total sensitivity index $S_T$. For the two 3N potentials $V_{\mathrm{3N}}^{(c_D)}$ and $V_{\mathrm{3N}}^{(c_E)}$, their subtracted matrices are 
\begin{equation}
    \overline{V^{(c_D)}_{3\mathrm{N}}} = \begin{bmatrix}
        0&2.73&0.09\\
        2.73&-1.21&0.15\\
        0.09&0.15&6.87
    \end{bmatrix}(\mathrm{MeV}),\quad \overline{V^{(c_E)}_{3\mathrm{N}}} = \begin{bmatrix}
        0&-0.02&-0.02\\
       -0.02&-3.38&-0.02\\
        -0.02&-0.02&-4.98
    \end{bmatrix}(\mathrm{MeV}).
\end{equation}
Compared to $V_{^3S_1}$, the impact of $V_{\mathrm{3N}}^{(c_E)}$ impact on the $^3S_1$-$^3D_1$ gap is milder, corresponding to a smaller total sensitivity index. In contrast, $V_{\mathrm{3N}}^{(c_D)}$ has a large non-diagonal matrix element, which would induce a large $^3S_1$-$^3D_1$ mixing. This constitutes the main reason why $M_{\mathrm{GT}}$ is sensitive to  $V_{\mathrm{3N}}^{(c_D)}$.

\subsection{S10. Convergence  check on the multi-channel variational method}

\label{sec:S8}

The multi-channel variational method amounts to a variational calculation within the subspace spanned by low-lying eigenstates of $H_S$. Therefore, the convergence of this method depends on the dimension of the subspace, i.e. the number of trial states $|\Psi^{i}_{T,\mathrm{C}/\mathrm{N}}\rangle$ used for calculation. In the following, we denote the set of labels of trial states  as $\boldsymbol{S}_\mathrm{C}$ and  $\boldsymbol{S}_\mathrm{N}$ for $^{14}$C and $^{14}$N respectively (see  Figure~\ref{fig:C14_trial_state} and~\ref{fig:N14_trial_state} for the labels). We start from the ground-state configuration of the shell model, successively including more configurations into the two set, and employ $H_{\chi}^{3\mathrm{N}_{\mathrm{opt}}}$ to perform variational calculation of $M_{\mathrm{GT}}$ within the corresponding subspace. In Figure~\ref{fig:PT_C}, we show the evolution of $M_{\mathrm{GT}}$ versus the change of the set $\boldsymbol{S}_{\mathrm{C}}$ for $^{14}$C. The number of trial states for $^{14}$N is fixed to be maximal, i.e. $\boldsymbol{S}_{\mathrm{N}}=\{b_1, b_2, b_3, b_4\}$.  It can be seen that $M_{\mathrm{GT}}$ shows a weak dependence on $\boldsymbol{S}_{\mathrm{C}}$, and convergence is achieved using only one or two $^{14}$C trial states.  In Figure~\ref{fig:PT_N}, we show the evolution of $M_{\mathrm{GT}}$ versus the change of the set $\boldsymbol{S}_{\mathrm{N}}$ for $^{14}$N. The number of trial states for $^{14}$C is fixed to be maximal, i.e. $\boldsymbol{S}_{\mathrm{C}}=\{a_1, a_2, a_3, a_4, a_5\}$. In contrast, $M_{\mathrm{GT}}$ is quite sensitive to the dimension of $\boldsymbol{S}_{\mathrm{N}}$. The convergence is not achieved until all $p$-shell configurations are included in the trial state set, corresponding to $\boldsymbol{S}_{\mathrm{N}}=\{b_1,b_2,b_3,b_4\}$. This is also a manifestation that $^{14}$C $\beta$ decay is sensitive to the shell structure of $^{14}$N. We also tested the effect of trial states with $sd$-shell valence excitation, finding their influence  negligible compared to statistical uncertainties.   

\begin{figure*}
\centering
\includegraphics[width=0.76\textwidth]{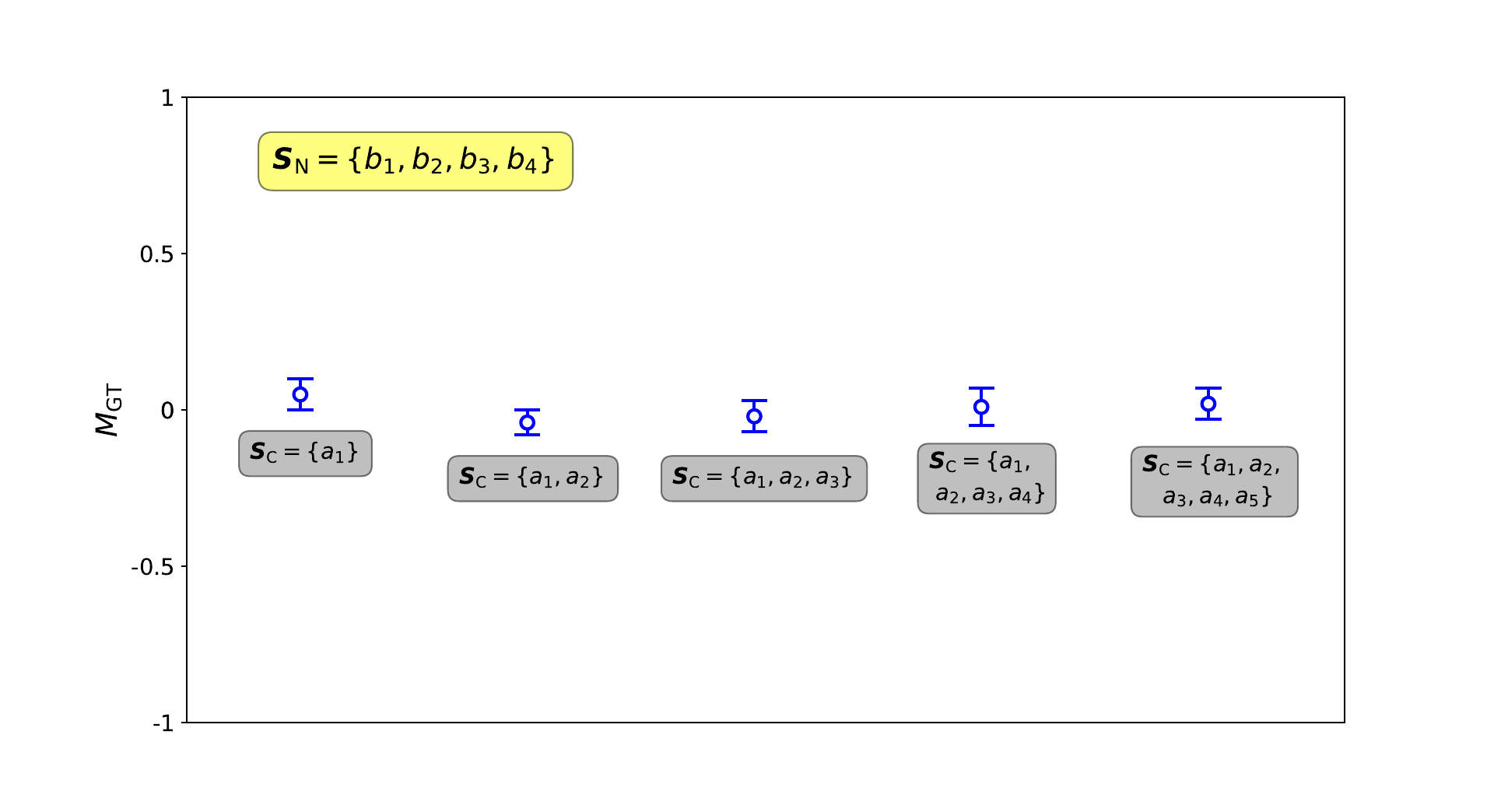}
\caption{The evolution of $M_{\mathrm{GT}}$ versus the change of $\boldsymbol{S}_{\mathrm{C}}$. The five points represent the results for $\boldsymbol{S}_{\mathrm{C}}=\{a_1\}, \{a_1,a_2\}, \{a_1,a_2,a_3\}, \{a_1,a_2, a_3,a_4\}$ and $\{a_1,a_2, a_3,a_4, a_5\}$ from left to right. The characters label the shell model trial states shown in Figure~\ref{fig:C14_trial_state}. All trial states in Figure~\ref{fig:N14_trial_state} are used for $^{14}$N, i.e. $\boldsymbol{S}_{\mathrm{N}}=\{b_1,b_2,b_3,b_4\}$. The error bar represents statistical uncertainty. }
\label{fig:PT_C}
\end{figure*}

\begin{figure*}
\centering
\includegraphics[width=0.76\textwidth]{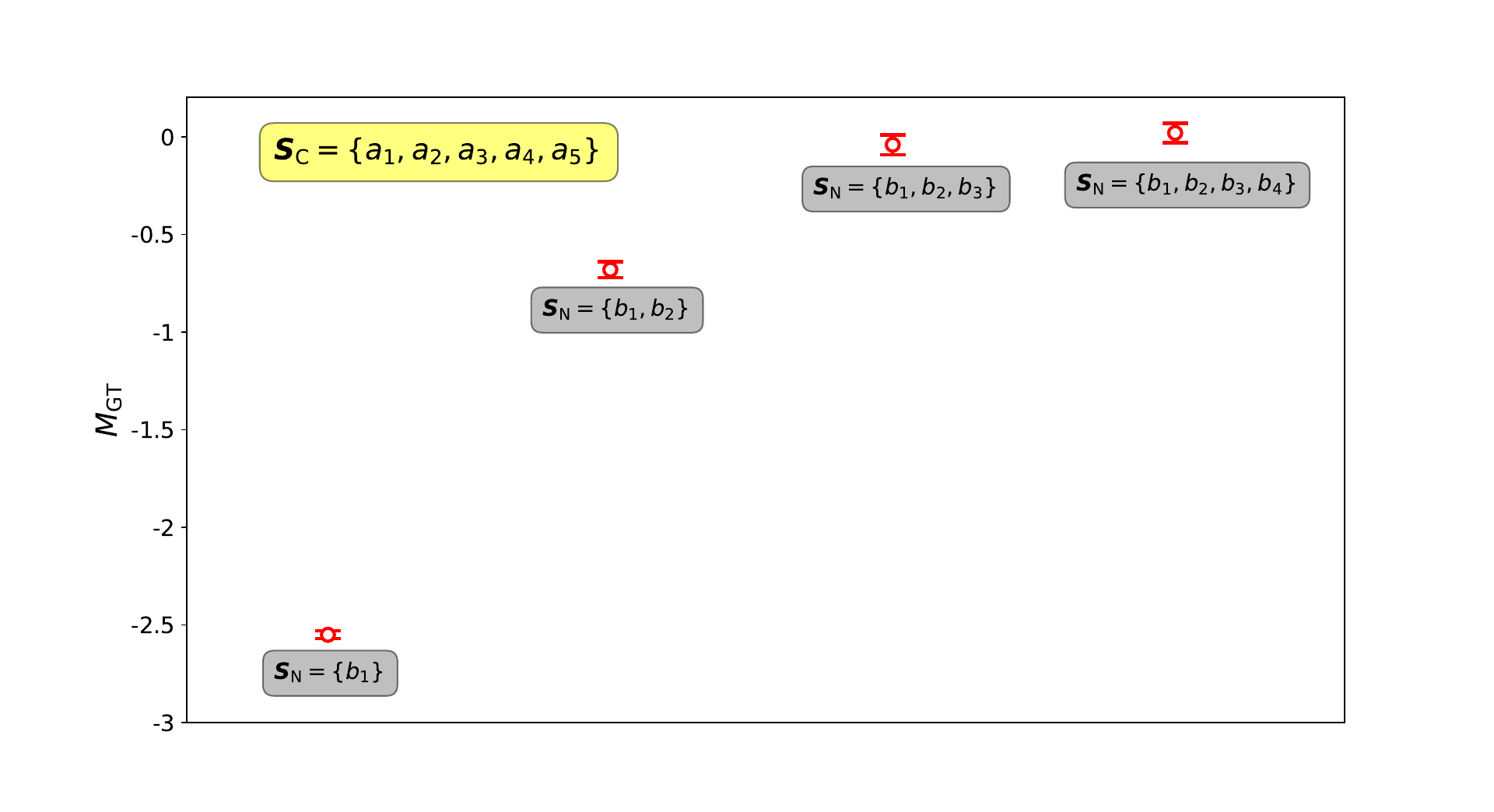}
\caption{The evolution of $M_{\mathrm{GT}}$ versus the change of $\boldsymbol{S}_{\mathrm{N}}$. The five points represent the results for $\boldsymbol{S}_{\mathrm{N}}$ = $ \{b_1\}, \{b_1,b_2\}$, $\{b_1,b_2,b_3\}$ and $\{b_1,b_2,$ $b_3,b_4\}$ from left to right. The characters label the shell model trial states shown in Figure~\ref{fig:N14_trial_state}. All trial states in Figure~\ref{fig:C14_trial_state} are used for $^{14}$C, i.e. $\boldsymbol{S}_{\mathrm{C}}=\{a_1,a_2, a_3,a_4, a_5\}$. The error bar represents statistical uncertainty.}
\label{fig:PT_N}
\end{figure*}

\subsection{S11. Details on the global sensitivity analysis}

\label{sec:S10}

To quantify the relative importance of the input LECs on the model output, Sobol's GSA is performed in this work. For more details, we refer the
reader to Ref. ~\cite{Sobol:2001, Saltelli:2002, Saltelli:2010}. Unlike a local sensitivity analysis, this variance-based approach explores the entire input space and accounts for nonlinearities and interactions between variables. The key idea is to decompose the variance ${\mathcal V}$ of the output into fractions that can be attributed to specific inputs and the  correlation between inputs, order by order,
\begin{equation}
    {\mathcal V} = \sum_{i} {\mathcal V}_i +\sum_{i,j}{\mathcal V}_{ij}+\sum_{i,j,k}{\mathcal V}_{ijk}\cdots,
\end{equation}
where $i$ labels the $i$th LEC $C_i$ in the Hamiltonian $H$. ${\mathcal V}_i$ is the  partial variance attributed solely to the uncertainty in $C_i$, and ${\mathcal V}_{ij}$ is the  partial variance due to the simultaneous variation of
$C_i$ and $C_j$, etc. The Sobol indices are computed
by dividing these partial variances by the total variance $\mathcal{V}$, such as the first-order indices (also known as the main effect) $S_i = {\mathcal V}_i/{\mathcal V}$,the second-order indices $S_{ij} = {\mathcal V}_{ij}/{\mathcal V}$, and so on. The total-order index $S_{Ti}$ (also known as the total effect) used in  Figure~\ref{fig:ST_distribution} is defined as
\begin{equation}
    S_{Ti} = S_i+\sum_{j}S_{ij}+\sum_{j,k}S_{ijk}+\cdots~.
\end{equation}
$S_{Ti}$ measures the total contribution of $C_i$ to the output variance, including its main effect and all higher-order interactions with other LECs. For practical simulations of a complex models, the sensitivity indices can be only calculated through Monte Carlo or Quasi-Monte Carlo sampling. For the GSA in this work, we consider  2N LECs up to NLO and all the 3N LECs, allowing them to vary in the range bounded by $10\%$ around their optimal values. We have also investigated the role of higher-order 2N forces and find their contribution to the total variance rather small, so we do not show them in the main text. Note that for the $c_D$ and $c_E$ terms in the 3N sector, only $V_{3\mathrm{N}}^{(c_D)}$ and $V_{3\mathrm{N}}^{(c_E)}$ of  Eq.~(\ref{eq:V_3N_in_short}) are varied independently, rather than all the 8 sub-terms of Eq.~(\ref{eq:cDcE_term}), i.e.
\begin{equation}  V_{3\mathrm{N}}^{(c_D)}\rightarrow \lambda_{c_D}V_{3\mathrm{N}}^{(c_D)},\quad  V_{3\mathrm{N}}^{(c_E)}\rightarrow \lambda_{c_E}V_{3\mathrm{N}}^{(c_E)}.
\end{equation} 
In this way, the number of 3N parameters entering the GSA is effectively reduced. Consequently, there are 14 independent parameters in total (9  LECs of the 2N contacts potentials, $c_1,c_3$ and $c_4$ in the 3N TPE potential as well $\lambda_{c_D}$ and $\lambda_{c_E}$), as is illustrated in Figure~\ref{fig:ST_distribution}. 
To allow fast calculations with controlled error rates, we follow  Saltelli's  scheme of  Sobol' sequence for sample generation, and we realize it through the corresponding functions provided by the
{\fontfamily{qcr}\selectfont python} library SALib~\cite{Herman2017, Iwanaga2022}. We  generate $(14+1)\times 2^{14}=245760$ samples in total, which enables the extraction of  statistically significant  total
effects. For each sample, we  calculate the corresponding values of $a_{L=2}$ and $M_{\mathrm{GT}}$, which are subsequently used as input for the calculation of   $S_i$ and $S_{Ti}$ using the {\fontfamily{qcr}\selectfont sobol.analyze} function in SALib.

\subsection{S12. The formula for the $^{14}$C lifetime} 

\label{sec:S11}

The lifetime $T_{1/2}$ of $^{14}$C shown in Figure~\ref{fig:C14_lifetime_contour} is calculated from the formula~\cite{Chou:1993}
\begin{equation}
    T_{1/2}=\frac{1}{f_A}\frac{K}{B_{\mathrm{GT}}},
\end{equation}
with $K=6139\mathrm{s}$ an overall constant~\cite{Dubbers:1991}, $f_A=10^{-2.208}$  the Gamow-Teller phase-space factor for $^{14}$C $\beta$ decay~\cite{Chou:1993} and $B_{\mathrm{GT}}$ the Gamow-Teller transition strength. $B_{\mathrm{GT}}$ is related to $M_{\mathrm{GT}}$ through
\begin{equation}
    B_{\mathrm{GT}} = \frac{{g}_A^2 M_{\mathrm{GT}}^2}{2J_i+1}.
\end{equation}
For the contour plot of $T_{1/2}$ versus $(\lambda_{c_D}, \lambda_{c_E})$ in Figure~\ref{fig:C14_lifetime_contour}, we only retain the central value of $T_{1/2}$ fitted from the lattice data and neglect the statistical uncertainty.

\end{onecolumngrid}
\end{appendix}

\end{document}